\documentclass[a4paper,11pt]{article}
\pdfoutput=1 

\usepackage{jheppub}
\usepackage{bm}
\usepackage[utf8]{inputenc}
\usepackage{xspace}
\usepackage{afterpage}

\newcommand{\mean}[1]{\ensuremath{\left\langle #1 \right\rangle}\xspace} 
\newcommand{\meanpt}{\ensuremath{\left\langle \pt \right\rangle}\xspace} 
\newcommand{\pt}{\ensuremath{{p_{\rm T}}}\xspace}
\newcommand{\pT}{\ensuremath{{p_{\rm T}}}\xspace}
\newcommand{\mt}{\ensuremath{{m_{\rm T}}}\xspace}
\newcommand{\piplus}{\ensuremath{\pi^{+}}\xspace}
\newcommand{\piminus}{\ensuremath{\pi^{-}}\xspace}
\newcommand{\kplus}{\ensuremath{K^{+}}\xspace}
\newcommand{\kminus}{\ensuremath{K^{-}}\xspace}

\newcommand{\pbar}{\ensuremath{\bar{\rm p}}\xspace}
\newcommand{\degree}[1]{\ensuremath{#1^{\circ}}\xspace}
\newcommand{\dbeta}{\ensuremath{1/\beta-1/\beta_{\rm calc}}\xspace}

\newcommand{\gevc}{\rm{GeV/$c$}\xspace}
\newcommand{\mevc}{\rm{MeV/$c$}\xspace}
\newcommand{\xF}{\ensuremath{x_{\rm F}}\xspace}
\newcommand{\pp}{\ensuremath{pp}\xspace}

\newcommand{\pythia}{\textsc{PYTHIA}\xspace}
\newcommand{\epos}{\textsc{EPOS}\xspace}
\newcommand{\sqrts}{\ensuremath{\sqrt{s}}\xspace}

\newcommand{\dndy}{\ensuremath{dN/dy}\xspace}

\title{\boldmath Production of identified charged hadrons vs.\ transverse momentum and rapidity in \pp collisions at $\sqrt{s} = 62.4$ and 200\,GeV}

\newcommand{\bnl}{Brookhaven National Laboratory, Upton, New York 11973, USA}
\newcommand{\krakow}{~Smoluchowski Institute of Physics, Jagiellonian University, PL-30348 Krakow, Poland}
 
 \newcommand{\nbi}{Niels Bohr Institute, University of Copenhagen, DK-2100 Copenhagen, Denmark}
 \newcommand{\texas}{Texas A$\&$M University, College Station, Texas 77843, USA}
 \newcommand{\bergen}{~University of Bergen, Department of Physics and Technology, N-5020 Bergen, Norway}
 \newcommand{\bucharest}{~University of Bucharest, RO-077125 Bucharest, Romania}
 \newcommand{\kansas}{~University of Kansas, Lawrence, Kansas 66045, USA}
 \newcommand{\oslo}{University of Oslo, Department of Physics, N-0316 Oslo, Norway}
\newcommand{\spacescience}{~Institute for Space Sciences, RO-077125 Bucharest, Romania}
 \newcommand{\iphc}{Institute Pluridisciplinaire Hubert Curien CNRS-IN2P3 et Universit{\'e} de Strasbourg, F-67037 Strasbourg, France}
\newcommand{\lund}{Div. of Experimental High-Energy Physics, Lund University, Lund, Sweden}

\author[j]{I.~Arsene,}
\author[d]{I.~G.~Bearden,}
\author[h,1]{C.~Besliu,\note{Deceased.}}
\author[d]{H.~B{\o}ggild,}
\author[a]{C.~Chasman,}
\author[a]{C.~H.~Christensen,}
\author[k]{P.~Christiansen,}
\author[d]{H.~Dalsgaard,}
\author[d]{J.~J.~Gaardh{\o}je,}
\author[f]{K.~Hagel,}
\author[d,1]{O.~Hansen,}
\author[h]{A.~Jipa,}
\author[d]{C.~E.~J{\o}rgensen,}
\author[c]{R.~Karabowicz,}
\author[a]{J.~H.~Lee,}
\author[c,1]{Z.~Majka,}
\author[i]{M.~Murray,}
\author[d]{B.~S.~Nielsen,}
\author[d]{C.~Nygaard,}
\author[b]{F.~Rami,}
\author[e]{C.~Ristea,}
\author[h]{O.~Ristea,}
\author[g]{D.~R{\"o}hrich,}
\author[i]{S.~J.~Sanders,}
\author[c]{P.~Staszel,}
\author[j]{T.~S.~Tveter,}
\author[a]{F.~Videb{\ae}k,}
\author[g]{Z.~Yin,}
\author[e]{and S.~Zgura}

\affiliation[a]{\bnl}
\affiliation[b]{\iphc}
\affiliation[c]{\krakow}
\affiliation[d]{\nbi}
\affiliation[e]{\spacescience}
\affiliation[f]{\texas}
\affiliation[g]{\bergen}
\affiliation[h]{\bucharest}
\affiliation[i]{\kansas}
\affiliation[j]{\oslo}
\affiliation[k]{\lund}

\collaboration{The BRAHMS Collaboration}

\emailAdd{videbaek@bnl.gov} 

\keywords{Hadron-Hadron Scattering, Heavy Ion Experiments, Forward Physics, Minimum Bias, Particle and Resonance Production}

\abstract{
The BRAHMS experiment at the Relativistic Heavy Ion Collider (RHIC) has measured charged pions, kaons and (anti)protons in \pp collisions at $\sqrt{s}= 62.4$ and 200\,GeV. 
Transverse momentum spectra have been measured for identified particles over a broad range in rapidity, from mid-rapidity, $y \sim 0$ to very forward rapidity, $y \sim 3-4$.
From the \pt spectra, the \pt integrated yields and the average \pt (\meanpt) are extracted and studied. Estimates of the 4$\pi$ integrated yields are also presented.
The net-proton (protons minus antiprotons) rapidity density distribution exhibits longitudinal scaling from SPS to RHIC energies indicating that there is no significant change in the mean rapidity loss in this energy range.
The observed long-range transport of baryon number over 5 units of rapidity can be described by a two-component model consistent with a contribution involving baryon junctions.
Finally, the results are compared with {\pythia}8 and \epos 1.99 model predictions. The rapidity and \pt spectra of pions and kaons are reasonably well described at both energies by {\pythia}, while \epos provides an overall better description, in particular of the proton and net-proton distributions.

}

\begin{document}
\maketitle
\flushbottom

\section{Introduction}

The scientific program of Brookhaven National Laboratory's Relativistic Heavy Ion Collider (RHIC) has produced measurements that indicate the formation of a
strongly-coupled Quark Gluon Plasma (sQGP) in colliding heavy ions
~\cite{BRAHMS:WhitePaper,PHOBOS:WhitePaper,STAR:WhitePaper,PHENIX:WhitePaper}, as well as new insights into the origin of the spin of the proton \cite{PhysRevLett.113.012001}.
Proton-proton (\pp) collisions form an important baseline that is imperative for a solid understanding of the heavy-ion results. 
There is also considerable current interest in understanding the large
transverse single spin asymmetries measured with pions and kaons in \pp collisions with center-of-mass energies
ranging from 20 to 200 GeV and at large Feynman x  (\xF)
~\cite{E704:adams91,E704:bravar96,BRAHMS:ANN_62gev,BRAHMS:ANN_Spin2006,STAR:abelev-2008-101,Adare:2013ekj,Adamczyk:2012xd}. Finally, we note that, as pointed out in Ref.~\cite{Pierog:2006qv}, the production of baryons and antibaryons in high-energy hadronic collisions is important for understanding cosmic-ray showers. For example,
the results presented here may shed light on the "muon puzzle"~\cite{EAS-MSU:2019kmv}.
Thus, we need to have both reliable measurements of identified hadron production over a wide rapidity and \pt range, as well as understand how these measurements are described by phenomenological models.

At RHIC, mid-rapidity particle production in \pp collisions has been addressed
in Refs.~\cite{Adare:2011vy,STAR:2003oii} (PHENIX, STAR) where the transverse
momentum distributions of pions, kaons, protons, and antiprotons
were measured. 
Forward particle production has been addressed in the
neutral pion channel~\cite{STAR:abelev-2008-101} and for selected rapidities
for charged pion, kaon, and (anti)proton production in Ref.~\cite{BRAHMS:pp2007PRL}.
Those data were used in Ref.~\cite{de_Florian_2007} in global fits for fragmentation functions for pions and kaons in hadron and electron colliding systems.
In this paper, we report for the first time the measurements of charged pions,
kaons, and (anti)protons in \pp\, collisions at $\sqrt{s}= 62.4$ and 200\,GeV
over a wide range in rapidity and with an extended \pt coverage, which allows  bulk quantities such as yields and average \pt (\meanpt) to be determined with good precision. These measurements constitute the final comprehensive results from the BRAHMS \pp data-taking program at RHIC.

The results are compared with two very different phenomenological models.
The \pythia model~\cite{PYTHIA,Sjostrand:2006za,Bierlich:2022pfr} describes inelastic non-diffractive \pp collisions as superpositions of parton-parton
collisions in which Lund string fragmentation~\cite{Andersson:1983ia} is used to hadronize the final state partons. 
The \epos model~\cite{Pierog:2009zt,Pierog:2013ria} has also been widely applied in describing heavy-ion data and cosmic-ray data. 
It includes physics that is similar to that of \pythia for the corona, but assumes that a QGP is formed in
the dense core-part of the collision, where hadronization is modeled via the hadron resonance gas model.
Both of the generators come in different variations.   
We have used {\pythia} 8.226 and \epos 1.99 throughout the paper, which we will just refer to as \pythia and \epos in the text. 

The paper is organized as follows:
Section \ref{sec:detector} discusses the BRAHMS detector system as it was configured for the \pp runs in 2005 and 2006. Additional details
are included for three sub-systems, which were not described in previous BRAHMS publications. 
The data sets collected are presented in Section~\ref{sec:dataset}.
Section ~\ref{sec:analysis} describes the data analysis, 
starting with a detailed description of the tracking algorithms used in both spectrometers, followed by the methodology used for the identification of the detected charged particles. 
Cross section extraction and the corrections applied to the data during that process are also 
described in this section, which ends with a summary of the estimated systematic uncertainties. 
Section~\ref{sec:results} is a thorough description of the results, starting with the transverse momentum distributions and particle
ratios. 
The obtained spectra are compared with existing measurements; for 62.4 GeV to results from the CERN 
Intersection Storage Ring (ISR), and for 200 GeV to RHIC data from STAR and PHENIX.
This section then provides rapidity distributions of the yields and the mean transverse momentum.
 Longitudinal scaling and the degree of stopping is investigated using the rapidity
distribution of net-protons, followed by a discussion of baryon transport modes. 
Finally, Section~\ref{sec:results} also presents the energy dependence of the average multiplicity and 
Section~\ref{sec:theory} describes the  comparison between the invariant yields extracted from \pp collisions at $\sqrt{s}=~62.4$ and 200 GeV with {\pythia}\ and \epos. 
 
\section{\label{sec:detector}BRAHMS Detector System}

The BRAHMS detector consists of two movable magnetic spectrometers: the Forward Spectrometer (FS) that can be rotated from
\degree{2.3} to \degree{15}, with a front part that can be rotated further up to \degree{30}, and the Mid-Rapidity Spectrometer (MRS) that can be rotated from \degree{34} to \degree{95} relative to the beam line, as described in 
Subsection~\ref{sec:spectrometers}. 
Three critical detector systems not covered
in previous BRAHMS publications, namely the vertex detector, the extended TOF in 
MRS and the spectrometer trigger system for pp running, are detailed in subsections~\ref{sec:vertex},
\ref{sec:TFW2} and~\ref{sec:triggers}.

\subsection{BRAHMS Spectrometers}
\label{sec:spectrometers}
The MRS is a single dipole-magnet spectrometer subtending a solid angle of ${\approx}5$\,msr and with a magnetic bending power of up to 1.2\,Tm. 
It has two time projection chambers (TPCs), TPM1
and TPM2, situated in field free regions in front of and behind the dipole magnet.
This assembly is complemented by two highly segmented scintillator time-of-flight walls, the first one referred to as TOFW, and the second one, TFW2.
The momentum resolution $\delta p/p$ lies between 2 and 8\%,  depending on momentum and magnetic field. 
The maximum \pt-reach of the hadron spectra in the MRS is 3--4\,\gevc.

The FS consists of 4 dipole magnets D1, D2, D3, and D4, with a total bending power of up to 9.2 Tm and a corresponding maximum momentum acceptance of ${\approx}40$\,\gevc.
The spectrometer has 5 tracking stations T1 through T5.
T1 and T2 are TPCs placed in front of and after the second dipole D2. 
T3, T4, and T5 are drift chambers each with a position resolution of ${\approx}80\mu\text{m} $ with
T3 in front of D3, T4 between D3 and D4, and T5 after D4.
T5 is followed by two particle identification detectors, a segmented time-of-flight wall (H2) and a ring imaging Cherenkov detector (RICH)~\cite{BRAHMS:RICH}. 
The D1-D4 magnets are all set to run at the same fraction of their full field value. 
Thus, the acceptance for a given magnetic setting covers a certain momentum
range around $p_{\rm ref}$, a reference momentum. 
At the highest field setting $p_{\rm ref}$ is 22\,\gevc. 
The momentum resolution is dominated by the position resolution of the
tracking detectors, and can be expressed as $\delta{p}/p = 0.016\,p/p_{\rm ref}$, implying a resolution between 0.8--2\% for accepted particles of interest at any given magnetic field setting.
Additional details on the BRAHMS experimental setup can be found in Ref.~\cite{BRAHMS:NIM} and in Ref.~\cite{BRAHMS:auau2005PRC} for tracking in the MRS. 

\subsection{Vertex and Luminosity detectors}
\label{sec:vertex}

A set of four Cherenkov Counters (CC) installed at 1.9 (inner ring) and 6.4 meters (outer ring) on either side of the nominal IP are used to measure the beam luminosity. 
The detectors are designed to achieve good time resolution, and are used to determine  the vertex of the collision and to provide the start time for all time-of-flight measurements.
Each detector consists of a 4.9 cm thick Lucite radiator backed by a small number of Photo-Multipliers Tubes (PMTs), with 8 or 5 PMTs (explained below) in the inner rings and 10 PMTs in the outer rings. 
The light collection in these detectors allows most of the Cherenkov light emitted by incident charged particles above the threshold to reach the PMT photo-cathodes.
The detectors cover the pseudo-rapidity range $3.26< |\eta| <5.15$.
All rings have full azimuth coverage, except the inner ring in the forward direction, which has a cutout for \degree{120}$ < \phi<$ \degree{240} to minimize material, and thus  production of secondary particles in front of the forward spectrometer.

An average timing signal is derived from all tubes hit in the CCs on either side of the IP.
The sum and the difference of these signals determine the start
time of the event and the vertex position of the interaction. 
From comparisons to vertices formed from tracks measured in the MRS spectrometer it is deduced that
the position resolution is ${\approx}1.2$\,cm, which corresponds to a time resolution of about 100\,ps. For further details see Ref.~\cite{BRAHMS:CC}

There are significant yields of charged particles in \pp collisions at 200 GeV within the rapidity  coverage of the CC detectors. 
The coincidence between the CC detectors serves as a minimum bias trigger and is used 
for luminosity determination.  
It is recorded with a typical scale-down  factor of 400.
The cross section for the CC  detectors is estimated from the Vernier scans  (described later) to be sensitive to ${\approx}68\%$ of the total inelastic cross section of 42 mb \cite{ParticleDataGroup:2018ovx}.
For \pp collisions at 62.4\,GeV the beam rapidity is smaller and the CC detectors are only sensitive to ${\approx}33\%$ of the total inelastic cross section of 36\,mb.

\subsection{TFW2}
\label{sec:TFW2}

The TFW2 time-of-flight detector is an array of 41 BC408 scintillator slats designed to measure the time-of-flight of charged particles
in the MRS. 
Each slat is 40\,cm high, 5\,cm wide and 1.5 cm thick, coupled with optical cement at both ends to R2083 PMTs (2 inch Hamamatsu H2431 assembly). 
The anode signal is passively split and one signal is fed to a FASTBUS ADC, while the other is connected to a discriminator for
timing purposes. 
The input signal to the discriminator is passed through a high frequency pass filter mounted at the tube base.
This detector is built to be symmetric about the MRS axis of rotation and can be placed at two radial positions at 5.59 m and 6.97 m 
measured from the pivot of the spectrometer to the center of the wall. 
The TFW2 is positioned at the larger distance for angles less than $60^o$. 
The scintillator slats are mounted on two arcs.
The front arc has a radius of curvature of 5.08\,m and the back of 5.12 \,m.
Both arcs are centered on the D5 magnet nominal center whenever the detector is at the shortest distance to the spectrometer pivot.
The overall time-resolution of this detector system is $\delta \text{t} \sim$120\,ps.

\subsection{\label{sec:triggers}Triggers}
The MRS trigger is formed by requiring coincidences between the TOFW placed at 4.33\,m from the IP,
a hodoscope array (TRMRS) placed immediately behind the D5 magnet, and the
RHIC 9.7 MHz clock. 
The TRMRS is a 12 slat scintillator hodoscope array, where each slat has
dimensions $2\times 9 \times 0.4$\,cm (W$\times$H$\times$D) and is read out by fast Hamamatsu H5010 phototubes at both ends of each slat. 
The slats were made thin to minimize the multiple scattering for low momentum particles. 
Details about the TOFW detector can be found in Ref.~\cite{BRAHMS:NIM}.
Analog signals from both TOFW and TRMRS detectors were fed into custom designed programmable modules.
These modules provide a logical AND from the top and bottom signals of each slat, and a logical OR from all slats.
The overlap coincidence time is 20\,ns. The trigger logic requires at least one good hit in each of the hodoscopes. 
The resolving time of the TOFW and TRMRS detectors is much smaller than the bunch crossing time of 107\,ns.
 
In the FS, the trigger is formed with signals from a hodoscope array (TRFS) placed immediately behind the D1 magnet and in front of the first tracking detector, T1, and from the two time-of-flight walls, H1 and H2, placed at 8.8\,m and 18.8\,m from the IP, respectively, as well as the RHIC 9.7\,MHz clock. 
The TRFS is a 7 slat hodoscope array with slat dimensions $3 \times 9 \times 0.4$ cm (W$\times$H$\times$D) readout by Hamamatsu H5010 phototubes at each end similar in design to the TRMRS.

The triggers for both spectrometers do not require the minimum bias CC trigger, and thus register tracks from events that are part
of the total inelastic \pp cross section, including single diffractive and double diffractive events. 
The efficiencies of both spectrometer triggers have been estimated using minimum bias data sets, and were found to be greater than 98\%.
The enhancement factor for these triggers are large: ${\approx}100$--$1000 $ depending on angle and field setting because of the small solid angle of the spectrometers. 

During the  200\,GeV \pp running period the largest event rates achieved for the FS were about 100 Hz,
which were handled by the data acquisition system with dead times $\le 25 \%$. 
For the MRS, the data were usually downscaled by factors of 3 to 5 in order to maintain good live time for the FS data. 
The dead time is dominated by the readout time of the TPCs.

\section{\label{sec:dataset}Data Sets}
The data used for this analysis were collected with the BRAHMS detector system during the 2005 and 2006 RHIC runs, which were the last BRAHMS data taking runs.
The energy of the 200 GeV \pp data matches previous heavy-ion runs (Au+Au and Cu+Cu collisions). 
The lower energy data (\sqrts = 62.4\,GeV) was collected during the short run in 2006.
The 62.4\,GeV energy was chosen to match the highest energy of the \pp collisions at the  ISR. 
The experiment sampled $0.26\,{\rm pb}^{-1}$ of \pp collisions at 62.4 GeV and
$3.4\,{\rm pb}^{-1}$ at 200 GeV.

The two BRAHMS spectrometers were run independently and the data are best characterized by the angular and magnetic field settings of each spectrometer.
The data taking conditions for 200 and 62.4\, GeV are summarized in Tables \ref{tab:settings62} and \ref{tab:settings200}, respectively.
These tables list the number of collected events at different spectrometer angles and magnetic field settings specified as the fraction  of the nominal maximum field. 
An effort was made to collect a similar number of events for both polarities of the magnets, and thus for positive and negative particles.
In the FS, positive (negative) particles are accepted for the A (B) polarity. 
Since the second goal of the experiment at both 200 and 62.4 GeV was to
measure transverse single-spin asymmetries for identified charged hadrons at
large values of the Feynman \xF in the forward region, the largest fraction of the beam time was spent taking data at \degree{3} and \degree{2.3} settings for the FS.

\begin{table}[htb]
  \centering
  \begin{tabular}{@{} |c|c|c|r|r| @{}}
    \hline
    Spectrometer& Angle & Field  & A Pol Trig & B Pol Trig \\
    \hline
    & degree & fraction &counts(k)&counts(k) \\
    \hline
    MRS
    & 90 & 0.16  & 810  & -\\
    & 90 & 0.31  & 1310 & 310 \\
    & 90 & 0.47  & -    & 90 \\
    & 45 & 0.31  & 2350 & - \\
    & 45 & 0.47  & 3490 & - \\
    \hline
    FS
    &   6 & 0.25  & 230 & 360 \\
    &   4 & 0.18  & 500  & 1100 \\
    &   4 & 0.50  & -    & 110 \\
    &   3 & 0.50  & 640  & 2020 \\
    & 2.3 & 0.50  & 1400 & 1030 \\
    \hline
  \end{tabular}
  \caption{
Angles, field settings and number of triggers in thousands (k)
for data taken at \sqrts = 62.4\,GeV. The field value is given as the fraction of the maximum for D1-D4 (9.2Tm) in FS, and D5 (1.6Tm) in MRS. In FS, A polarity is for positive particles and B for negative particles.
     }
  \label{tab:settings62}
  \end{table}

\begin{table}[htb]
  \centering
  \begin{tabular}{@{} |c|c|c|r|r| @{}}
    \hline
    Spectrometer& Angle & Field  & A Pol Trig & B Pol Trig \\ 
    \hline
        & degree & fraction &counts(k)&counts(k) \\
    \hline
    MRS
    & 90 & 0.16  & 1390 & -\\
    & 90 & 0.31  & 690  & - \\
    & 90 & 0.47  & 3510 & 970 \\
    & 90 & 1.00  & 2890 & 5760 \\
    & 60 & 0.31  & 260  & 320 \\
    & 45 & 0.31  & 700  & - \\
    & 45 & 0.47  & 6140 & 7830 \\
    & 40 & 0.31  & 490  & 170 \\
    & 34 & 0.31  & 1930 & 1250 \\
    & 34 & 1.00  & 5870 & 3140 \\
    \hline
    FS
    &   8 & 0.18  & 140   & 370 \\
    &   8 & 0.35  & 30    & 60 \\
    &   8 & 0.50  & 30    & 60 \\
    &   8 & 0.71  & 30    & 30 \\
    &   4 & 0.12  & 190   & 220 \\
    &   4 & 0.25  & 430   & 520 \\
    &   4 & 0.35  & 230   & 60 \\
    &   4 & 0.50  & 300   & 520 \\
    &   4 & 0.71  & 360   & 190 \\
    &   4 & 1.00  & 1080  & 1230 \\
    & 2.3 & 0.25  & 720   & 1660\\
    & 2.3 & 0.50  & 680   & 550\\
    & 2.3 & 1.00  & 15220 & 33030\\
    \hline
  \end{tabular}
  \caption{
    Angles, field settings and number of triggers in thousands (k)
    for data taken at \sqrts = 200\,GeV. The field value is given as the fraction of the maximum for D1-D4 (9.2Tm) in FS, and D5 (1.6Tm) in MRS. In FS, A polarity is for positive particles and B for negative particles. 
  }
  \label{tab:settings200}
\end{table}

\section{\label{sec:analysis}Data analysis} 
\subsection{Tracking}
\label{sec:tracking}

Local tracks are first determined as straight segments in the TPCs and Drift Chambers, which are all situated outside the magnetic field regions. 
The resulting straight-line tracks from two tracking chambers located on either side of a
magnet are matched using the effective edge approximation. 
The momentum $p$ of a particle with charge $q$ is determined by:
\begin{equation}
p/q = \frac{B l }{(sin(\phi_b)-sin(\phi_f)) \sqrt{1-\alpha_y^{2}}},
\end{equation}
where $B$ is the magnetic field, $l$ the length of the track from the forward effective edge to the backward magnetic effective edge, $\phi_f$ the angle between the track and the normal to the effective edge in front of the 
magnet at the position of the effective edge, $\phi_b$ is the same quantity at the backward effective edge, 
and $\alpha_y$ the average of the vertical slope of the track.
The magnetic field inside the D4 magnet gap has a spatial non-uniformity that requires a second correction to the deduced momentum.
This correction depends on the specific orbit of the track and it was deduced from full GEANT simulations of the spectrometer using a field map generated by the TOSCA program matching the measured D4 field. 
Straight-line track segments reconstructed locally in each tracking detector are extrapolated through the magnetic field using the effective edge approximation, and matched across field free regions to build complete tracks in the FS spectrometer.
Complete tracks are refitted to determine a single final value of $p/q$. 
The tracks in the FS are also required to project through the D1 magnet onto the nominal beamline to be accepted.

A number of track quality cuts are applied to select good tracks. 
The magnitude of related corrections and the evaluation of systematic uncertainties arising from them are discussed later in Section~\ref{sec:systematic}.

The selection criteria are as follows:
\begin{itemize}
\item Local tracks are matched between the tracking chambers, i.e., TPCs and Drift Chambers, to form complete tracks.
For matching tracks from each side of a dipole  magnet, three cuts are applied to track parameters based on the effective edge approximation, namely
the horizontal angle difference in the bending plane, the vertical slope,  and the position difference in the non-bending vertical plane.
Whenever tracks reconstructed in the T2 TPC and the T3 drift chamber are matched, the absence of magnetic field 
between those detectors calls for a different approach and a six-$\sigma$
elliptic cut in the differences of the  projected positions
and slopes on the matching plane
is applied.
The mean (offset) and resolution ($\sigma$) of the distributions used in the track matching are determined from data on a setting-by-setting basis.

\item Fully reconstructed tracks are extrapolated back to the primary vertex.
The distance of closest approach of the track and the beam line
 is compared with the z (longitudinal) coordinate of the vertex position, which was measured by the CC detectors. 
In the absence of a CC vertex, the track is projected to the nominal beam
intersection point, and required to be within three $\sigma$ in the vertical direction.

\item Fiducial cuts are made to require the tracks to clear the vertical and horizontal boundaries of the magnets by more than 1 cm.

\item A fiducial cut is made on the last PID detector. 
For the FS the thin walled window with dimensions $40 \times 20$ cm$^2$ in front of the RICH detector must be intersected.
 For the FFS it is required that the track intercepts an active H1 slat. 
For MRS, the track must intercept the active TFW2 slats.

\item For particle identification using time-of-flight, tracks are matched to hits in the 
TOF walls. 
A track is accepted if the slat it projects to or its immediate neighbor  has a signal. 
A 3$\sigma$ match in the Y position is also required. The Y coordinate of a hit is determined from the time difference between signals from the corresponding top and bottom PMTs. 
\end{itemize}

Tracking and matching efficiencies for each of the five tracking stations in the forward spectrometer are calculated by constructing full tracks using just
4 track segments and evaluating the efficiency in the $5^{th}$ station by comparing the predicted position and direction of the interpolated or extrapolated 
full track in that station with the known local segments.
The local track efficiency for each tracking station as a function of position and direction of the track segments is evaluated at each spectrometer angle and field setting.
The overall tracking efficiency is about 70-80\%, and is included in the calculation of the cross sections and per event yields.
The systematic uncertainty on the tracking efficiency is about 2--4\%.

\subsection{Particle identification}
\label{sec:pid}

In the MRS, particle identification is done using the time-of-flight with
the CC providing the start time and the TFW2 (or TOFW) the stop time. The TOFW
time-of-flight is used to cross-check the results from the TFW2. 
As a consequence of the longer flight path, the momentum range for good particle identification can be extended whenever the TFW2 is used at the cost of a small reduction in raw yield caused by the additional decay and absorption of particles.

\begin{figure}[htb]
\centering
 {\includegraphics[width=3.2in]{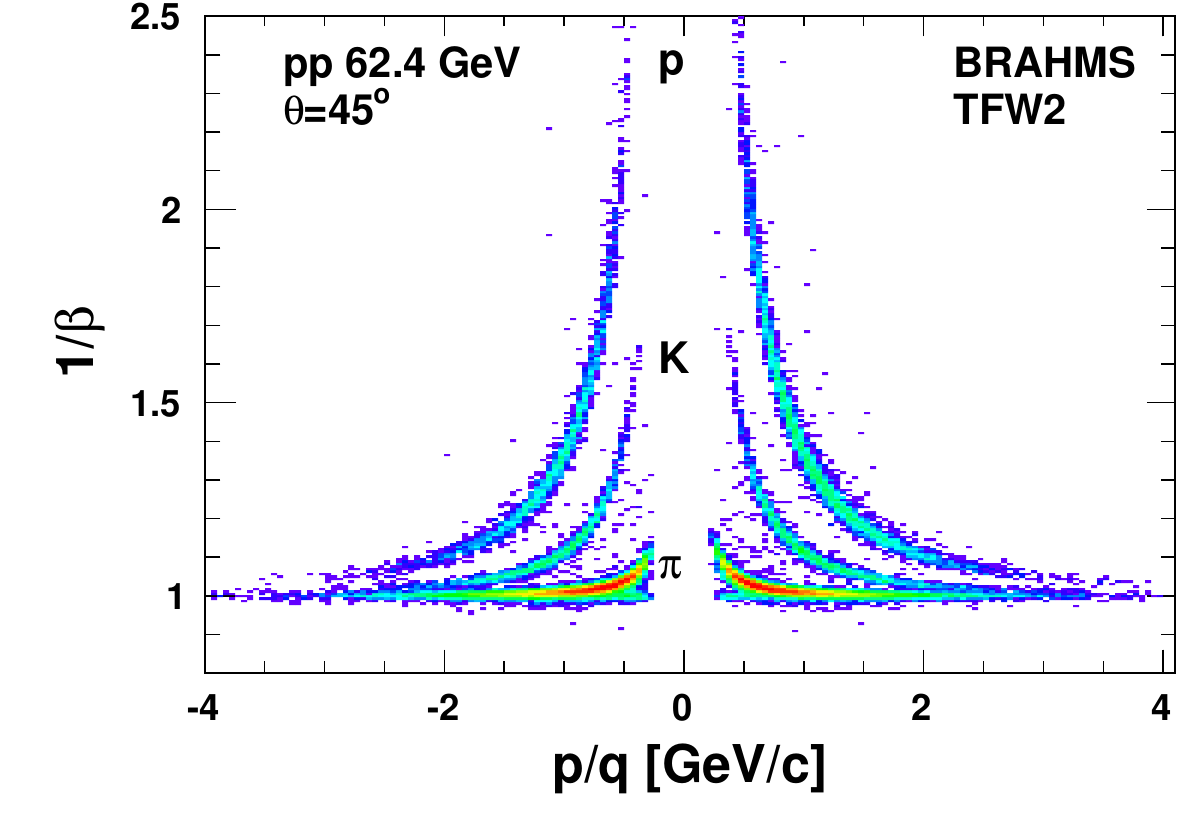}} 
 \caption{\label{fig:pidbeta45B1050}
 Distribution of $1/\beta$ (as measured by TFW2) vs. p/q for $\sqrts = 62.4$ GeV. Bands for pions, kaons and protons  are clearly seen.
 }
\end{figure} 

\begin{figure}

\centering
 \includegraphics[width=3.2in]{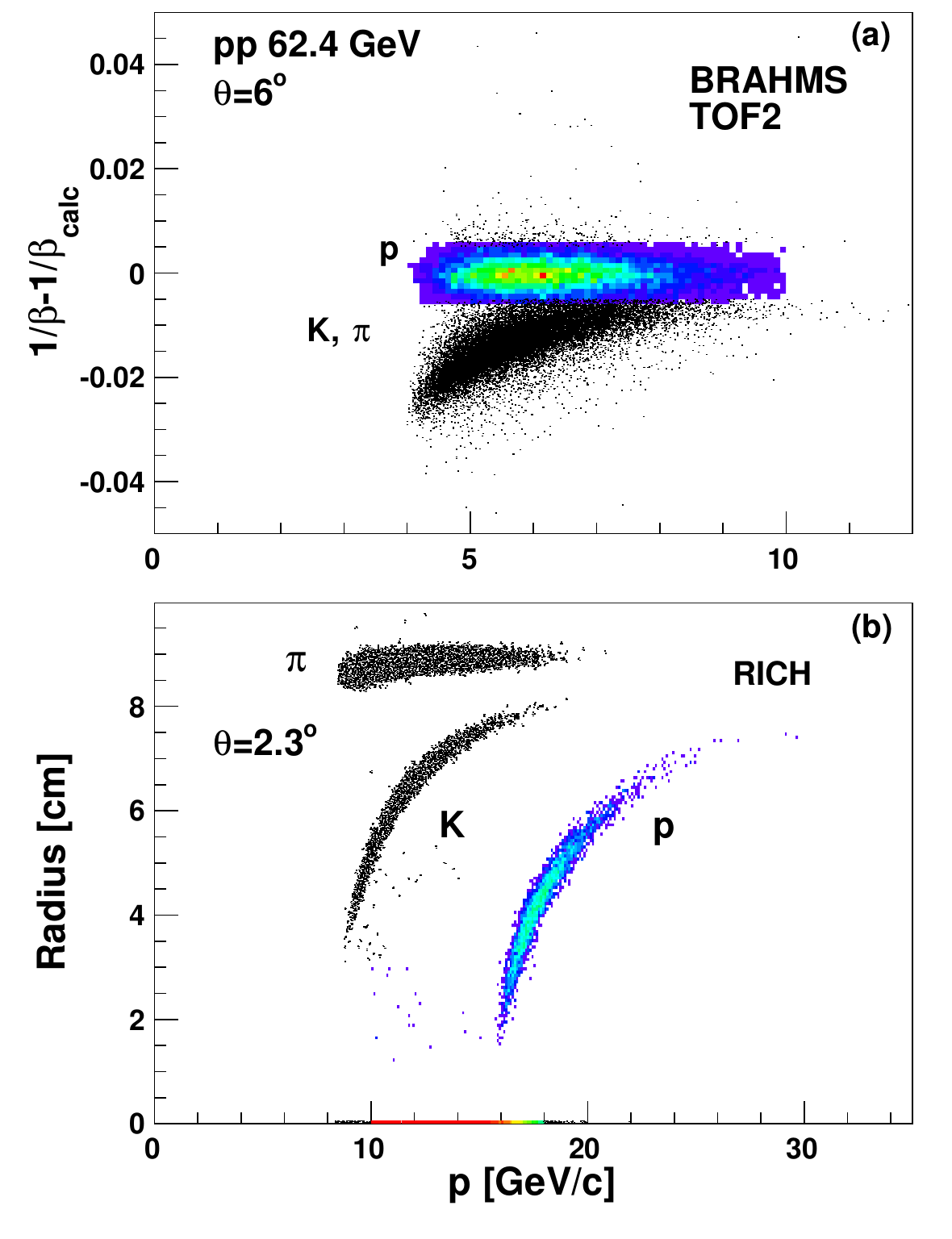}

\caption{\label{fig:LFPID}
 Example of proton identification in the FS. Top panel: proton
 identification using time-of-flight at \degree{6}.
 Bottom panel: proton identification using RICH at \degree{3} (half
 field). In both cases the particles identified as protons are highlighted in
 color. Note that this includes a large group of protons identified using the
 VETO mode (neither pions nor kaons) in the bottom panel (Radius $= 0$\,cm).
 }
\end{figure}

To identify charged pions, kaons and protons with the time-of-flight
detectors, a three standard deviation $\sigma$ cut in $1/\beta-1/\beta_{calc}$
is applied, where $\beta_{calc} = |p|/\sqrt{p^2+m^2}$ is the calculated velocity given the mass hypothesis $m$, and $\beta$ the measured velocity.
Figure~\ref{fig:pidbeta45B1050} shows a typical correlation between velocity and momentum of charged particles detected in the MRS spectrometer at $45^{\circ}$, demonstrating the good overall particle identification.
To derive the PID from time-of-flight, an event vertex is needed to calculate the path length and time signals from both sides of the CC counters are needed to calculate the start time.  
These requirements, which are not always fulfilled, have important consequences for the normalization as discussed later.
The resolution in $1/\beta$ for TFW2 has a typical value of 0.0055 at $34^{\circ}$ to $60^{\circ}$ and 0.007 at $90^{\circ}$.
Kaons are well separated from pions up to 1.8\,\gevc, while protons are separated from kaons up to 3\,\gevc. 
The pion spectra can be extended to somewhat higher momenta because the $K/\pi$ ratio is well below 1 (${\approx}$0.2-0.35 increasing with \pt). 
The quantity \dbeta was sampled in momentum slices for momenta above 1.8\,\gevc with bins of $\delta \pt$ of 0.2\,\gevc to 
extract the purity of kaons and pions by fitting a double Gaussian distribution to the spectra.
This was used to correct the yields at \degree{90}\ and \degree{45}.
Below 1.8\,\gevc the contamination of pions with kaons is negligible, but typically grows to approximately $30\%$ for \piplus at 2.6\,\gevc and to around $25\%$ for \piminus at 2.7\,\gevc.
The  spectra of pions from the \degree{90} and \degree{45} extend up to \pt $=2.8$\,\gevc  and \pt$=2.2$\,\gevc, respectively.

In the FS, particle identification is made primarily with the RICH detector \cite{BRAHMS:RICH}, and with the H1 and H2 time-of-flight walls at the lower magnetic field settings. 
For all angle and field settings used, the pions are above the threshold in the RICH detector, and are identified by requiring the measured ring radius to be within $3\sigma$ of the calculated radius on a track-by-track basis. 
The efficiency of the RICH detector has been studied both with pions identified with the H2 time-of-flight detector in an overlapping momentum range and with a detailed GEANT-based simulation of the RICH detector, as described in Ref.~\cite{BRAHMS:RICH}.
The efficiency rises rapidly with momentum and reaches 97\% at momenta 20\% above threshold.
The yields are corrected by the RICH efficiency.
Kaons are identified in the momentum range $10<p<20$\,\gevc\ using the same technique, with the additional requirement that the measured radius is more than $3\sigma$ away from the pion radius at a given momentum.

Protons and antiprotons are identified using the RICH in veto mode for the momentum range $10<p<18$\,\gevc, where protons will not produce a signal in the RICH; most pions and kaons in this momentum range will emit Cherenkov light, but a small fraction, typically ${\approx}2.8\%$, will not leave a signal. 
This inefficiency is caused by secondary interactions and scatterings after the H2 hodoscope and before the active RICH volume and its magnitude is well reproduced by Monte Carlo simulations.
The contribution from these events, mimicking protons identified using the VETO
mode, is subtracted on a statistical basis from the (anti)proton
candidate spectrum under the assumption that they constitute
$2.8\%$ of the measured pions and kaons in the same setting and momentum range.
For protons this correction is negligible at all angle settings.
At the highest rapidity for antiprotons at 62.4 GeV the contamination constitutes about $80 \%$ of the yield of the antiproton candidates and results in an estimated  systematic uncertainty of $50 \%$. 
At the next highest rapidity (3.0) the systematic uncertainty is ${\approx}20\%$. 
For the 200 GeV data the uncertainty due to this contamination is less than $10 \%$.

In the momentum range $3<p<8$\,\gevc\,, i.e., for the \degree{4}-\degree{8} 
settings, protons and kaons are identified by using the time-of-flight in H2 and requiring that no signal is observed in the RICH detector, thereby eliminating the pion signal.
The yields of the protons can be estimated from the TOF data.
The kaon contribution to the protons, per momentum bin, is determined from fitting the timing distribution with multiple Gaussian functions and subtracting this contamination of the protons in the final analysis.
The purity of protons is 0.8 to 1.0 in this momentum range.
The yield of kaons is determined in the momentum range $3<p<6$\,\gevc\.from the TOF, where p/K identification is clean and where pions leave signals in the RICH. There is a further correction to the Kaon yield from misidentified pion of the order 12-20$\%$.
The quality of the PID separation of protons from kaons and pions is illustrated in Fig.~\ref{fig:LFPID}.

\subsection{Cross Section determination}
\label{sec:crosssection}

The differential cross section for hadron production is calculated from the minimum bias data, or from the triggered data set where
the minimum bias condition may or may not be required, as

\begin{equation}
E \frac{d^3\sigma}{dp^3}
= \mathcal{L}^{-1} { \frac{N_{\rm h}(y,\pt)} 
{ 2\pi \pt \,f_{\rm h}(y,\pt) \, {A_c \epsilon_{rec}} \,                                {\Delta \pt \Delta y} }}
\label{eq:sigma}
\end{equation}
where $\mathcal{L}$ is the integrated luminosity for a particular data set, $\rm{N}_h$ the number of counts in a given 
$\Delta \pt \Delta y$ wide bin at a given \pt and y value, and $f_{\rm h}$ is the fraction of the inclusive hadron yield where the minimum bias condition is satisfied.
The acceptance correction is given by $A_c$ and all the efficiencies associated with the hadron tracking and particle identification are included in the $\epsilon_{rec}$ factor. 
For the triggered data set, which does not require the CC condition, the factor $f_{\rm h}$ is 1. 
The following section is devoted to a discussion of the determination of each of these factors.

\begin{figure}[htb]
\centering
 {\includegraphics[width=3.2in]{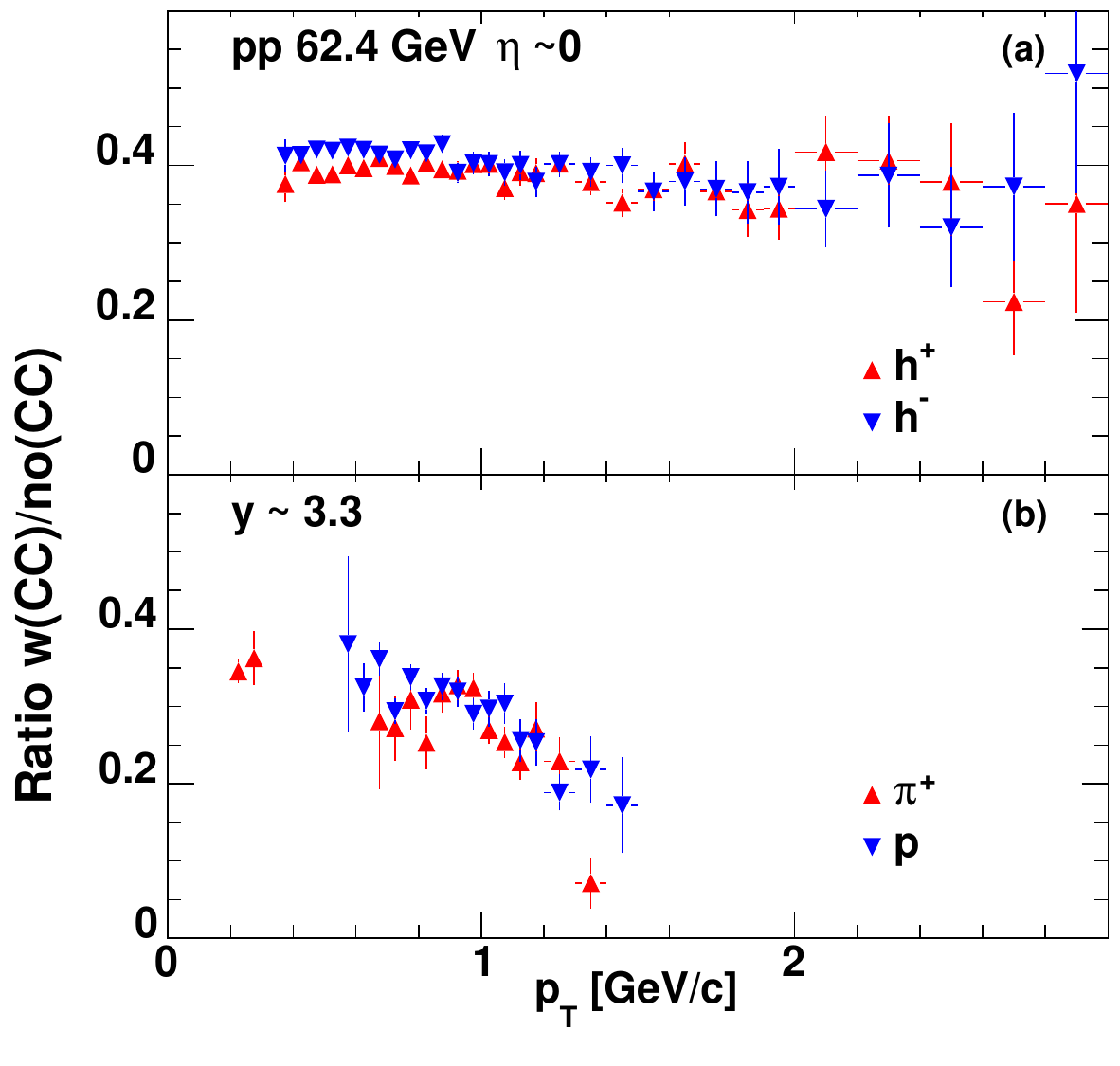}}
 \caption{\label{fig:ccrat_62} 
 Ratio of charged-hadron cross section at mid-rapidity for event classes with and without a CC-vertex for 62.4 GeV (a)  and for identified pions and protons near rapidity 3.3 (b).
 }
\end{figure} 

\begin{figure}[htb]
\centering
 {\includegraphics[width=3.2in]{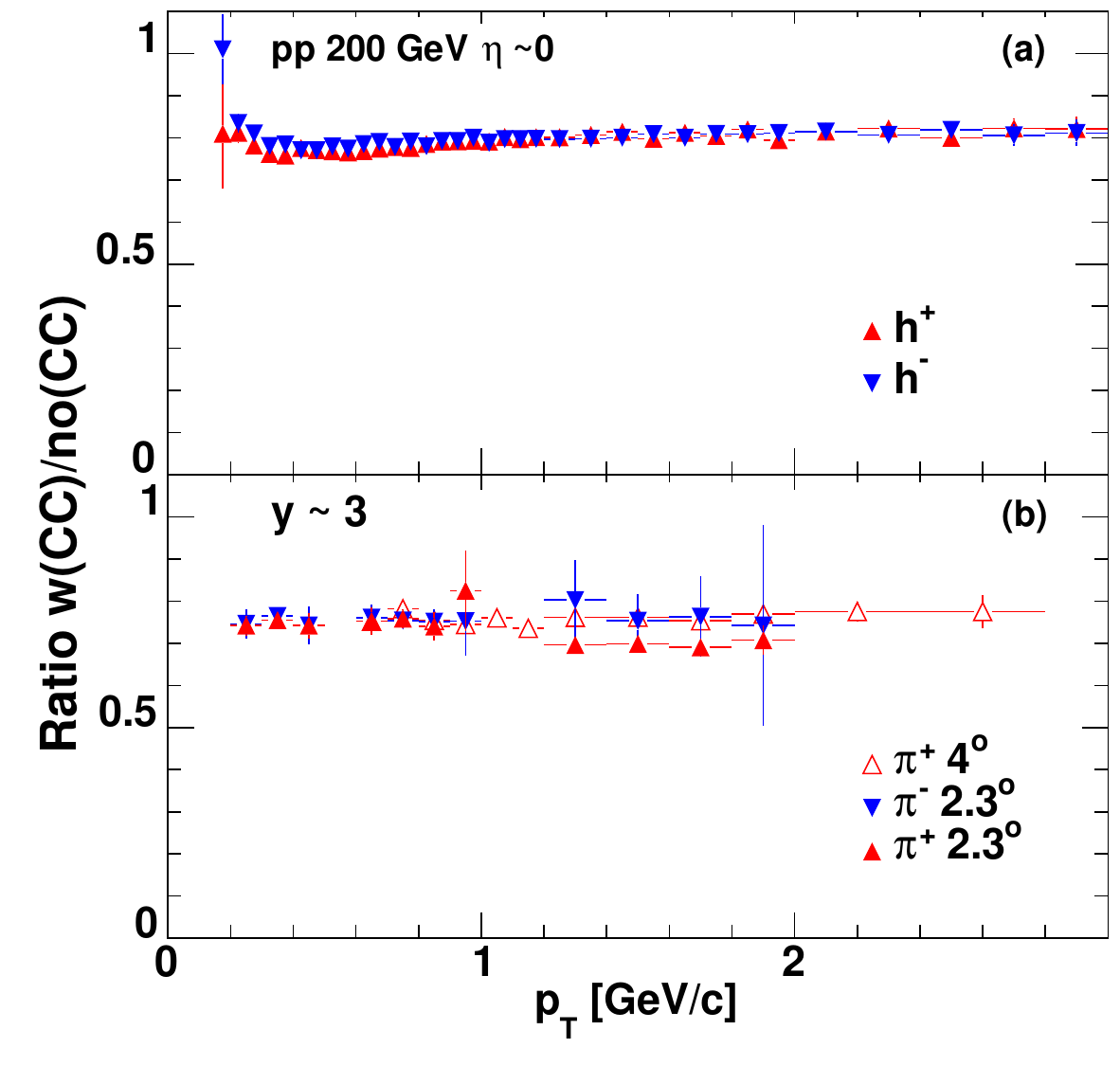}}
 \caption{\label{fig:ccrat_200} Ratio of charged-hadron cross sections at mid-rapidity  for event classes with and without a CC-vertex at 200 GeV  (a) and for identified pions near rapidity 3 (b).
 }
\end{figure}

\subsection{Luminosity determination, normalization and trigger bias}

The luminosity is deduced from the measured counts with the CC counters from minimum bias data, $ N_{cc} =\mathcal{L} \sigma_{\text{CC}}$.
The minimum bias CC trigger requires a minimum of one hit in each side of the CC detector system. 
The accepted vertex range measured with these detectors covers the full interaction region from
$-150$ to $+150$\,cm. 
The $\sigma_{\text{CC}}$ was evaluated using the method of Vernier scans~\cite{vernierscan2003}.
Details for the scans performed at 62.4 and 200 GeV data are given below.
It is noted that the systematic uncertainties in the normalization between the 62.4 and 200 GeV data are mainly uncorrelated since they are derived from two independent measurements.

\subsubsection{\rm{62.4 GeV}}

The $\sigma_{\text{CC}}$ was evaluated from two separate Vernier scans and is found to have a value of $12.0\pm 1.4$ mb. 
This corresponds to about $40\%$ of the Non Single Diffractive (NSD) cross section of 27\,mb and ${\approx}33\%$ of the total inelastic cross section of 36\,mb. 
A full GEANT simulation was performed of the response of the CC counters to  inelastic \pythia events, and  good agreement was found between the measured  and the simulated (13.3\,mb) fractional cross section, as well as to the hit distributions in the different CC counter rings.

Because the minimum bias CC trigger has such a small efficiency, it introduces
a bias towards selecting events with a high multiplicity of particles when the global vertex is required for event selection. 
This bias is not present for those spectrometer triggers that do not require the vertex information.
A correction for this bias is applied where the PID is derived from time-of-flight measurement where the start time is obtained from the CC coincidence data.
The bias, when present, may have a \pt-dependence that is evaluated for the forward spectrometer settings from the data using RICH
information only by comparing \pt spectra with and without the requirement of a global vertex. 
For the MRS, it was evaluated from $h^+$ and $h^-$ spectra with and without the
vertex requirement (see Fig.~\ref{fig:ccrat_62} (a)) and the \pt-dependence is
found to be less than $10\%$ at mid-rapidity. 
This dependence was also checked for pions with $\pt < 2.0$\,\gevc using the MRS trigger counter (TRMRS) as the start detector.
 In the FS, there is little dependence for \pt$<1$\,\gevc, which covers the majority of settings where time-of-flight is used.
 For the higher \pt region the PID is done using the RICH detector and no CC-vertex is required.
The effect is demonstrated in Fig.~\ref{fig:ccrat_62}(b) where we plot the ratio of yields obtained with and without a CC vertex requirement, i.e., the hadron fraction $f_{\rm h}$ needed for the cross section calculation. 
In the bottom panel the ratio for pions and protons  at forward rapidities is shown.
Note that here the reduction at large \pt is clearly due to exhausting the available energy for particle production. 
Therefore $f_{\rm h}$ is taken to be constant for \pt\ less than 1.0\,\gevc . This correction is only applied for data that requires time-of-flight information in FS where all \pt are less than 1~\gevc.

\subsubsection{\rm{200 GeV}}

For the 200 GeV data, the analysis requires that each event had a CC vertex associated with it, both at mid-rapidity and forward rapidity.
Therefore the experimentally measured cross sections are the invariant yields for the Non Single Diffraction (NSD). 
The inelastic cross section is obtained using Eq.~\ref{eq:sigma} with the additional knowledge of the $\sigma_{CC}$ and the factor $f_{\rm h}$.
Unfortunately, we do not have a very precise Vernier scan measurement at this energy, but only a value of $\sigma_{\text{CC}} \sim 28.0 \pm 3.5 $ mb, i.e.,  a $13\%$ uncertainty. 
The efficiency of the CC counter was cross-checked by Monte Carlo simulation using inelastic \pythia\ events as input as done for the 62.4 GeV data. 
This resulted in an estimated cross section of $\sim 27.5 \pm 2.0$ mb, depending on the tunes selected, consistent with the Vernier scan measurement.
Figure \ref{fig:ccrat_200} shows the ratio $f_{\rm h}$ for the 200 GeV data at
$y \sim 0$ (a) and at high rapidity (b). 
Below a \pt value of 3\,\gevc, no \pt-dependence is observed for the ratio. This is consistent with the observation by PHENIX for $\pi^0$ production at mid-rapidity~\cite{Adler:2003pb} where no \pt-dependence was observed for the corresponding ratio up to 10\,\gevc. 
 
\subsection{\label{sec:Corrections}Further corrections}

The data are corrected for the geometrical acceptance of the spectrometers, multiple scattering, weak decays of pions and kaons, and absorption in the material along the path of the detected particles.
It is assumed that the geometric acceptance and the corrections due to the different physical processes that particles are subject to can to first order be factorized. 
This procedure was confirmed by full Monte Carlo simulations, using as input single particle distributions that have spectral shapes similar to the observed ones, and found to reproduces the input spectra to better than $2\%$ overall. 
An exception is observed at mid-rapidity where deviations are seen for protons with momenta less than 0.7\,\gevc, kaons less than 0.5\,\gevc, and pions less than 0.4\,\gevc.
An additional correction is therefore applied for the low momentum mid-rapidity data.
This effect is primarily caused by a reduction of yield resulting from  multiple scattering out of the MRS acceptance at low momenta,  not compensated by scattering into this acceptance.
In the forward spectrometer no such additional correction is needed. 
\begin{figure}[tph]
\begin{center}
\begin{minipage}[b]{0.48\textwidth}
\centering
 \includegraphics[width=3.11in]{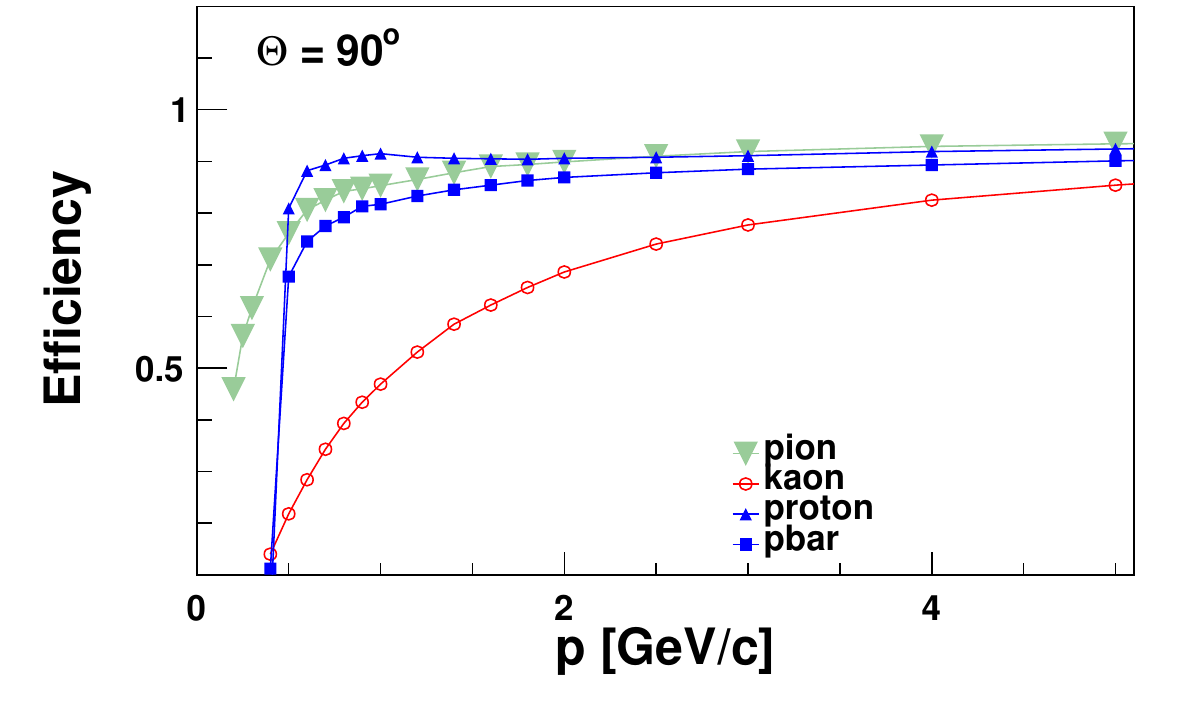}
\end{minipage}%
\end{center}
\begin{center}
\begin{minipage}[b]{0.48\textwidth}
\centering
 \includegraphics[width=3.11in]{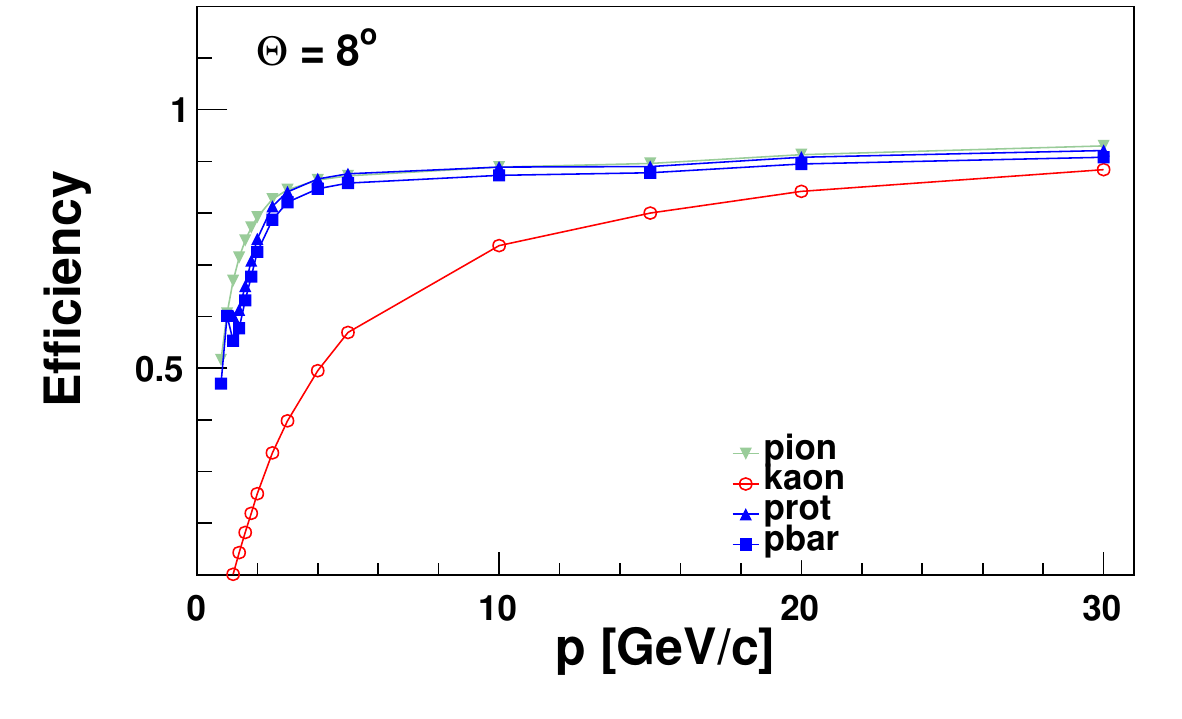}
\end{minipage}
\end{center}
\begin{center}
\begin{minipage}[b]{0.48\textwidth}
\centering
 \includegraphics[width=3.11in]{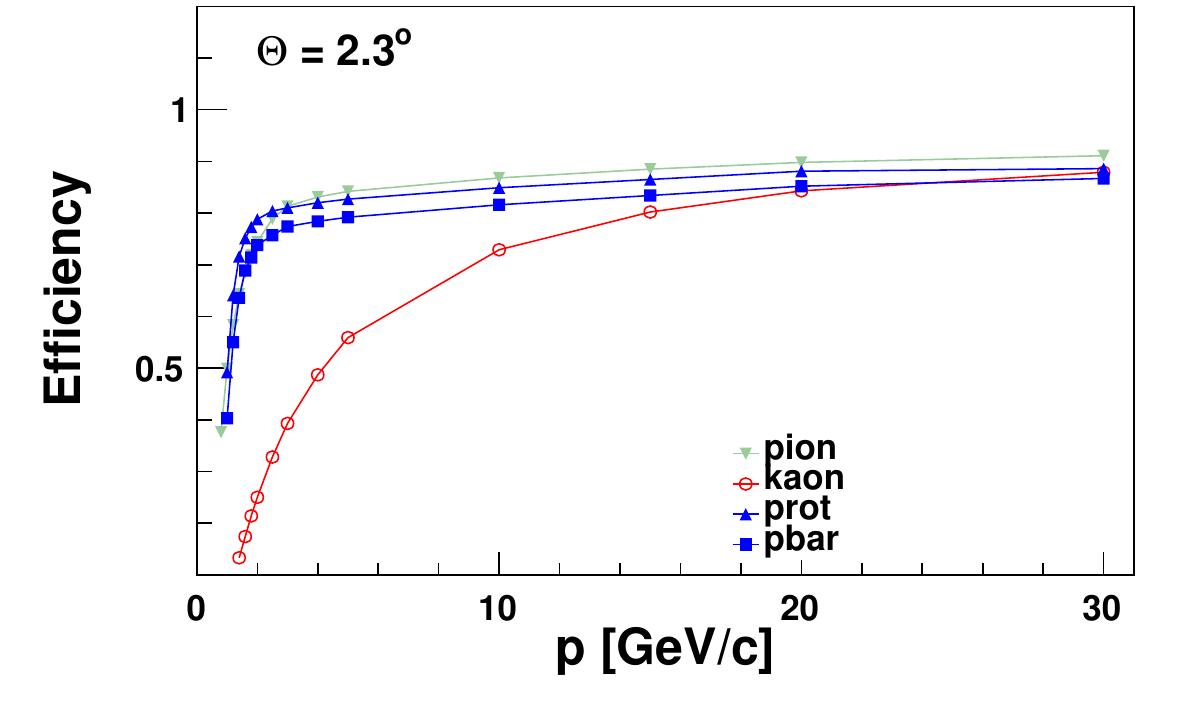}
\end{minipage}
\end{center}
 \caption{\label{fig:correction_90deg} 
 Reconstruction efficiency for pions, kaons and protons in the
 MRS (a) at \degree{90}, in the FS at \degree{8} (b), and  \degree{2.3} (c). 
 }
\end{figure}

The BRAHMS spectrometers are small solid-angle devices, so the largest correction to the recorded yield is from the geometrical acceptance of the spectrometers. 
It is evaluated by a purely geometric Monte Carlo with a detector description 
equivalent to what is used for the more sophisticated analysis based on detailed and complete GEANT simulations.
Particles are thrown from different vertex positions along the beam line and sorted into 5 cm wide vertex bins.
The probability that the particles traverse the spectrometers
at any given field setting, and within the relevant fiducial volumes, i.e., the TFW2 wall, the RICH detector, or the H2 hodoscope, are recorded.
For each particle type, a record is kept of this geometric probability as a function of y and \pt. 
The vertex bin size is slightly larger than the position resolution of the CC counters
and the deduced cross sections are found to be insensitive to the use of a smaller bin size.
The accuracy of the full acceptance correction is better than $1.5\%$, so even though the correction is large --
about a factor of 50-200 -- it is well determined.
 
The correction due to the interactions and decays that particles experience in the spectrometers is evaluated as follows.
For each kind of particle, pions, kaons, protons, and antiprotons, the correction is determined as a function of momentum and spectrometer angle setting using a dedicated MC setup based on the GEANT3 libraries~\cite{Brun:1987ma} with a realistic description of the BRAHMS detector system.
The default physics modes, parameters, and cuts are used. 
GEANT3 has two different packages for evaluating hadronic interactions.
The hadronic interactions are evaluated with the GEANT-FLUKA interface~\cite{GEANTFLUKA}. 
For \pbar\ at low momenta there is a significant difference compared with the
default GHESIA interface, while corrections for other particles are
within 1\% using the two different packages. 
Comparison of these codes to \pbar-absorption data shows that GEANT-FLUKA provides a
better description of this annihilation process at low momenta than the default GHESIA hadronic description in GEANT3.
Single particles are propagated through the spectrometers, the hits in the detectors are digitized and subjected to the same reconstruction and analysis package as real data.
The corrections are calculated for all angular settings used.
The accuracy of these corrections is estimated to be ${\approx}1\%$.
In Fig.~\ref{fig:correction_90deg} we show this correction as a function of momentum and particle type at \degree{90}, \degree{8}, and \degree{2.3}.
The dominant loss in the forward spectrometer is from hadronic interactions in the spectrometer path including the Be beampipe. 
The efficiencies of different particle species are  angle independent within $5\%$ for the FS.

\subsection{Vertex dependencies}

The differential invariant cross 
sections depend on the number of counts in a y-\pt bin, $N(y, \pt)$, as seen from Eq.~\ref{eq:sigma}.
Since the vertex distribution of pp collisions is rather wide ($\sigma_{z} \sim$ 60 cm) 
and the spectrometer acceptance depends on the vertex position, most strongly for the MRS, 
data from different vertex bins have to be merged, done by replacing Eq.~\ref{eq:sigma} with
\begin{equation}
 Ed^{3}\sigma/dp^{3} = \\ \mathcal{L}^{-1} \frac{\sum_v N_h(y,\pt,v)}{2\pi \pt\ \Delta \pt \Delta y\,\sum_v \, f_{\rm h}(y,p_t)\, A_c(v) \,\epsilon_{rec} }.
\label{eq:vtxsigma}
\end{equation}
Effects due to low statistics have to be taken into account.
This summing procedure preserves the proper Poisson statistics and ensures correct treatment of bins located within the acceptance but having zero counts, predominantly found at larger values of \pt.
A detailed discussion of this procedure can be found in Ref.~\cite{brahms:pcthesis}.

As a consequence of the small acceptance of the FS, a number of y-\pt~bins lie on the edge of this acceptance. 
We reject bins where the acceptance is less than 60\% of the average acceptance of bins in the rapidity range for a given setting. 
The data are analyzed in smaller y-\pt~bins (typically bins of $0.05 \times 50$\,\mevc) while results are presented for larger bins. 
In cases where a larger bin is not fully populated 
by smaller bins, the yield is therefore corrected based on interpolation, using the mean 
local slope of the \pt-spectrum for a particular rapidity bin.
At the most forward rapidities and the smallest field settings this correction can be up to 15\%. 
The correction due to the variation in cross section with rapidity is estimated to be less than a few percent, and is considered in the systematic uncertainty.

\begin{table*}[ht]
  \centering
  \begin{tabular}{|l|c|c|c|}
\hline
&Typical &Systematic&From\\
&value  &uncertainty & \\
&& &\\
 \hline 
CC Cross section 62 GeV& 12.0 mb & 1.4 mb & Vernier scan \\
\hline 
CC Cross section 200 GeV& 28.0 mb & 3.5 mb & Vernier scan \\
\hline
$f_h$ for MRS 200 GeV&0.75&0.02& data\\
\hline
$f_h$ for MRS 62 GeV &0.40&0.02& data\\
\hline
$f_h$ for FS 200 GeV&0.76&0.03& data\\
\hline
$f_h$ for FS 62 GeV &0.35&0.03& data\\
 
\hline
Trigger Efficiencies&0.99&0.02&data\\
\hline
Tracking efficiency (FS) &0.73&4\%&determined\\
     &   & &from data \\
\hline
Tracking efficiency (MRS) &0.90&4\%&determined\\
     &   & &from data and simulations\\
\hline

PID-RICH&0.97&$1\%$&data and\\
  & & &simulations \\
\hline
PID-TOF&0.98-0.80&$2-5\%$&from data\\

\hline
GEANT Corrections&0.6-0.85& $2\%$& simulations\\
\hline
Geometric Acceptance&0.02& $1\%$ &simulations\\
\hline
\end{tabular}

\caption{\label{tab:systuncertainties}
Correction factors and their systematic uncertainties. See discussion in text for details.
}
\end{table*}

\subsection{Systematic uncertainties}
\label{sec:systematic}

Table~\ref{tab:systuncertainties} summarizes corrections applied to the spectra and presents typical values and their systematic 
uncertainties, including how they were evaluated. 
The tracking efficiencies cover the track finding efficiency for individual detectors as well as track matching between detectors. The GEANT corrections contain losses due to nuclear interactions, multiple scattering, and weak decays of pions and kaons. The acceptance correction is different from the tracking efficiency in that it is purely geometrical (for last defining detector element). 
The PID efficiency contains both matching of tracks to PID detector signals as well as the selection efficiency.

The systematic uncertainties are largely independent of \pT. 
This means that the main effect is to change the overall normalization and therefore most of the systematic uncertainties cancel in \pt-dependent and integrated ratios. 
Similarly, all corrections except the PID and GEANT ones are independent of particle species and therefore mostly cancel in ratios, e.g., $K/\pi$. 
For \pt spectra, typical values of the total systematic uncertainty are ${\approx}15\%$ at $y \sim 0$ and  ${\approx}16\%$ at $y \sim 3$. For antiparticle-to-particle ratios, the total systematic uncertainty is small, e.g., of order $4\%$ at $y \sim 0$ and $5\%$ at $y \sim 3$. See also Section~\ref{sec:particleratio}. 
For ratios between different hadron species, typical values of the total systematic uncertainty are ${\approx}5\%$ at $y \sim 0$ and ${\approx}6\%$ at $y \sim 3$.  (See section \ref{sec:densities}).
For most systematic uncertainties there is little difference between mid-rapidity and forward data.
For the yields, \dndy, there is an additional contribution from the \pT extrapolation that will be discussed later.

\afterpage{\clearpage}

\section{\label{sec:results}Results}

\begin{figure*}[htb]
\centering
 {\includegraphics[width=0.49\textwidth]{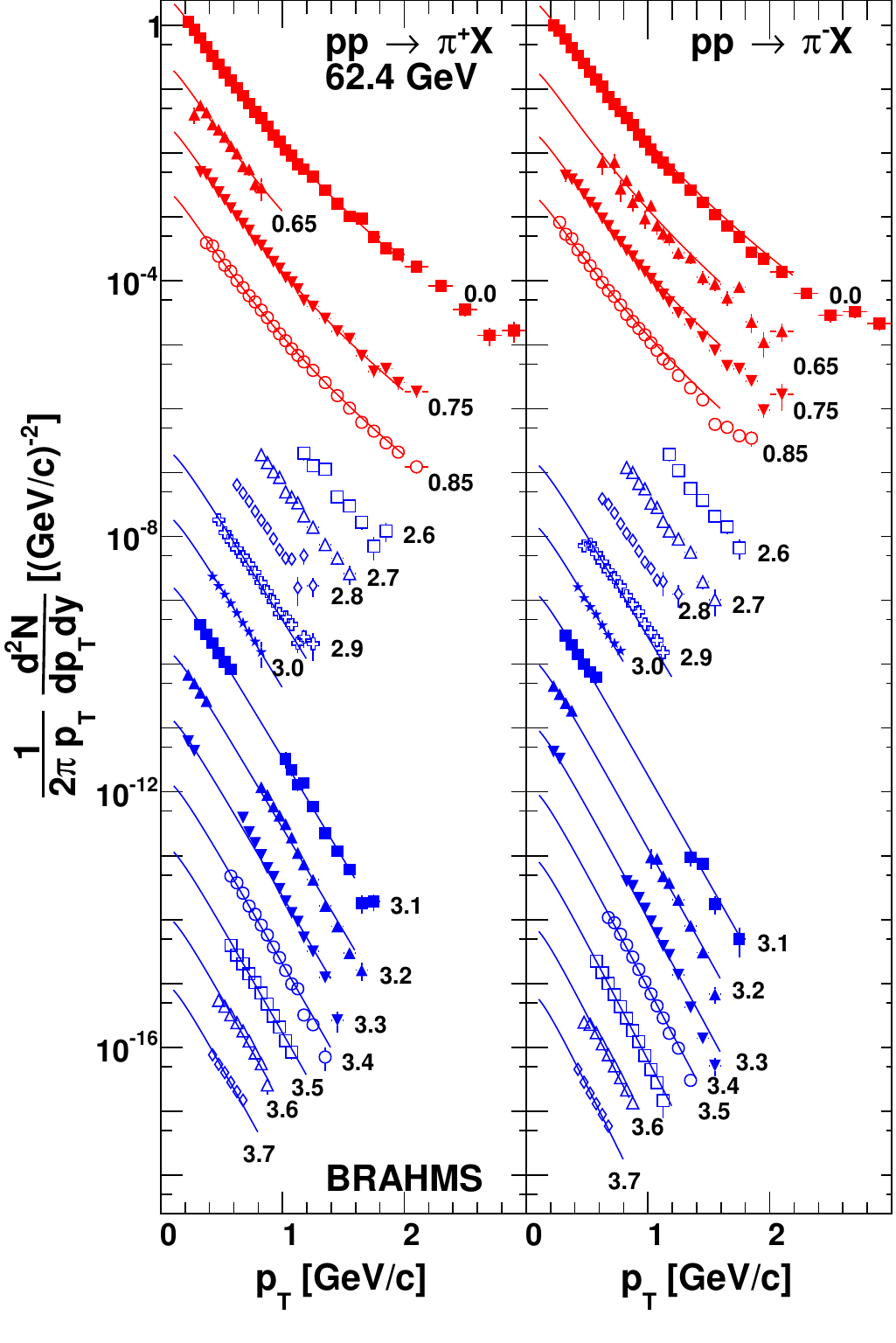}}
  {\includegraphics[width=0.49\textwidth]{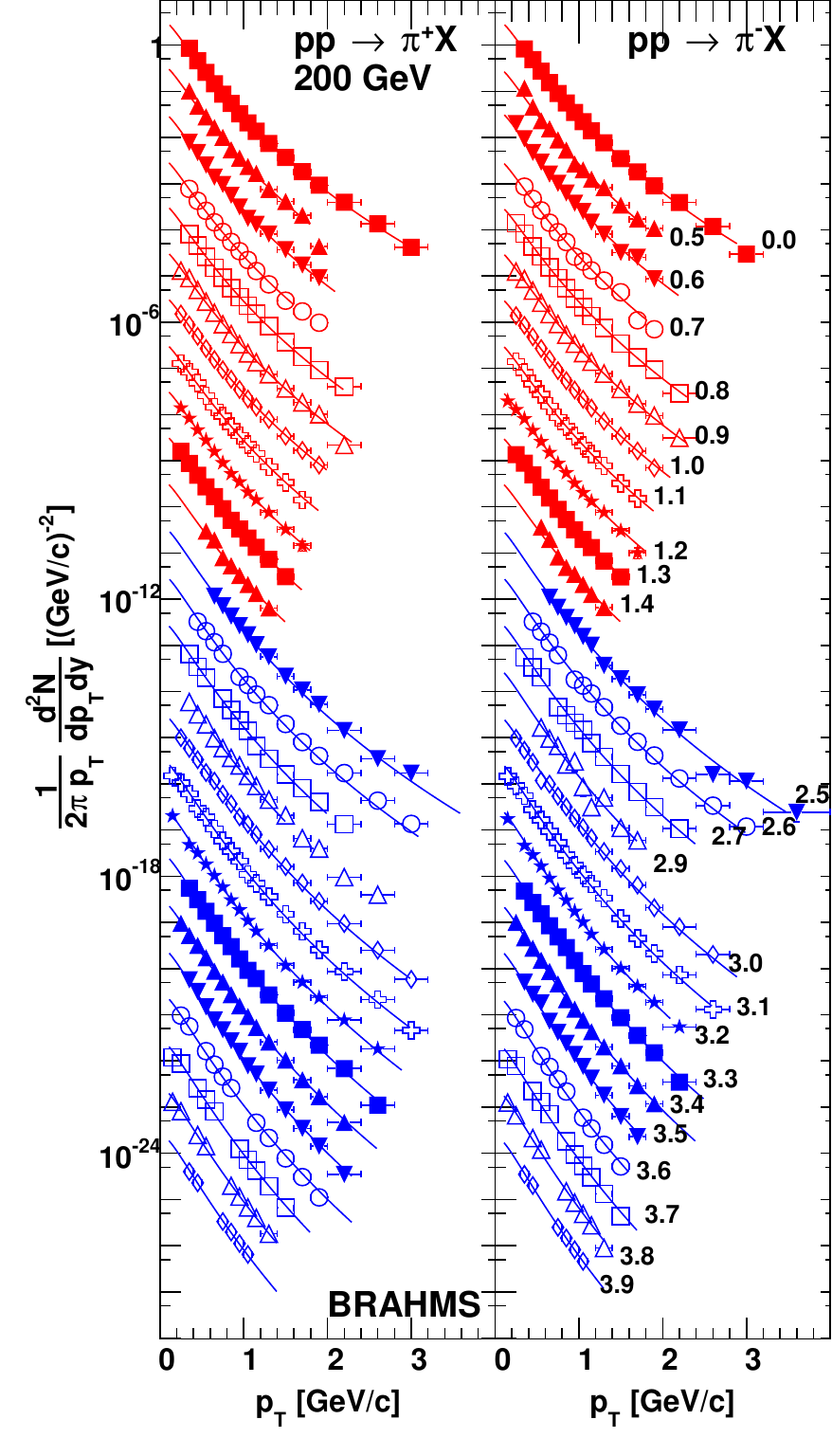}}
 \caption{\label{fig:pionspectra} Invariant yields per event as a function of the transverse momentum \pt\ for pions measured in \pp collisions at 62.4 and 200 GeV (two left most and two right most panels, respectively). 
 For each energy \piplus is shown to the left, and \piminus to the right.
 Rapidities are indicated at the end of each spectrum in \pp collisions at 62.4 and 200 GeV.
 Each rapidity bin is scaled down by a factor of 10 from the previous one. 
 The spectra for rapidities below 1.4 shown on top were measured with the MRS spectrometer (red), the rest were measured with the FS spectrometer (blue).
 The curves are the fits used to extract the integrated yields, see text for details.
 The statistical uncertainty is indicated by vertical bars. For clarity, the systematic uncertainties are not shown, see Section~\ref{sec:systematic} for typical values.
 }
 
\end{figure*}

\begin{figure*}[tb]
\centering
 \includegraphics[width=0.49\textwidth]{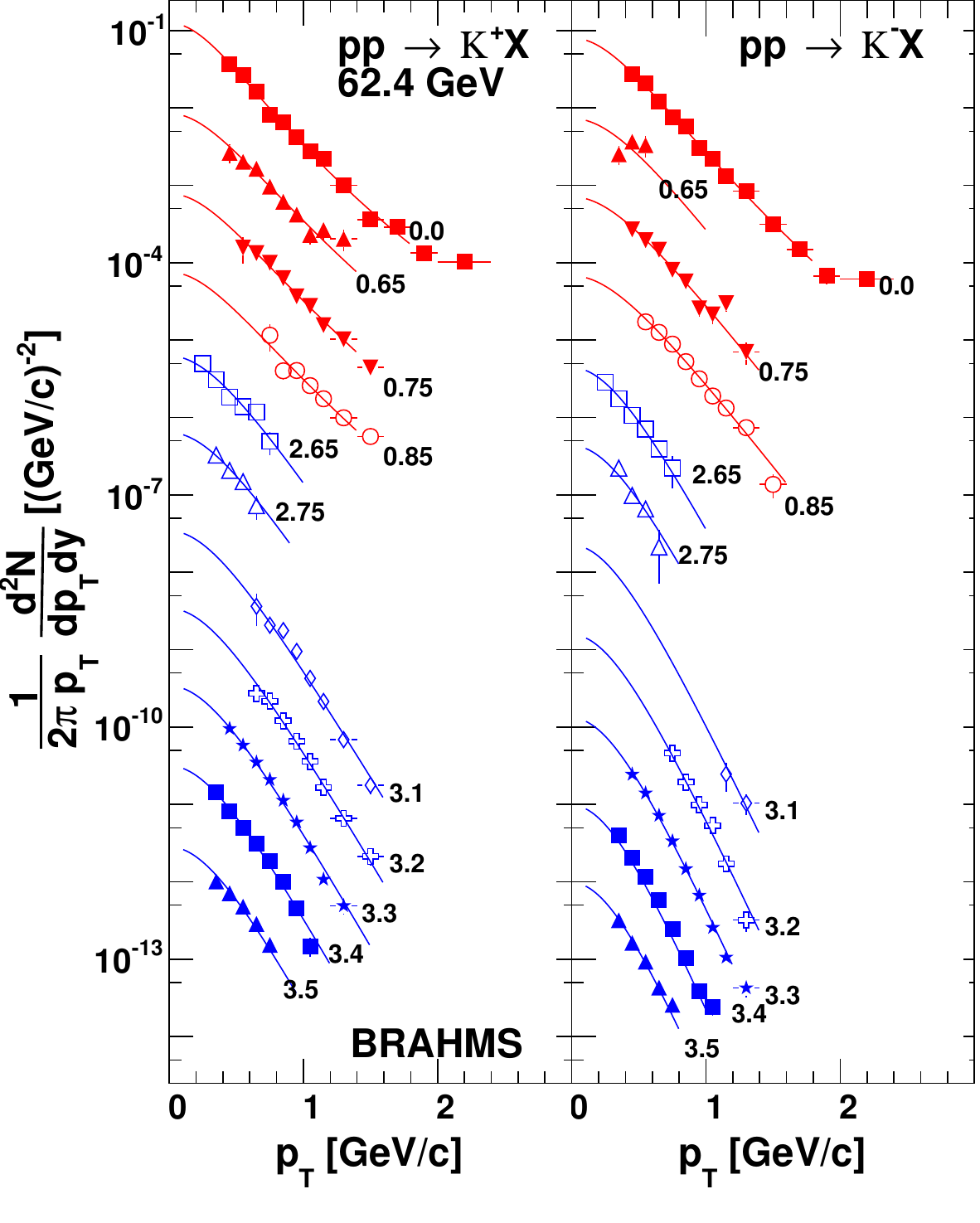}
  {\includegraphics[width=0.49\textwidth]{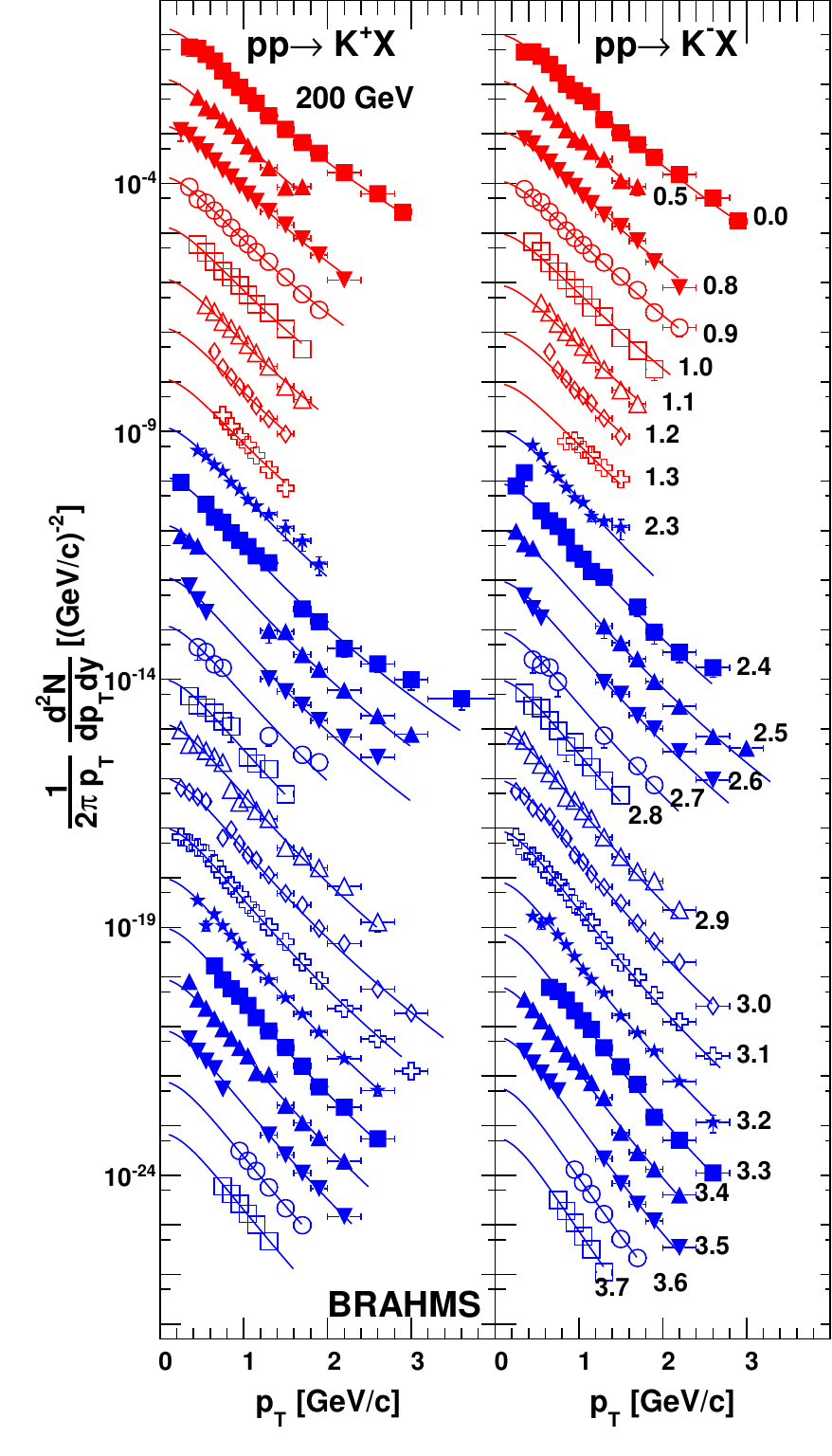}}
 \caption{\label{fig:kaonspectra} Invariant yields per event as a function of the transverse momentum \pt\ for kaons  measured in \pp collisions at 62.4 and 200 GeV (two leftmost and two right most panels, respectively). 
 For each energy \kplus is shown to the left, and \kminus to the right.
Rapidities are indicated at the end of each spectrum in \pp collisions at 62.4 and 200 GeV.
 Each rapidity bin is scaled down by a factor of 10 from the
 previous one. 
 The spectra for rapidities below 1.4 shown on top were measured with the MRS spectrometer (red), the rest were measured with the FS spectrometer (blue). The curves are the fits used to extract the integrated yields, see text for details.
 The statistical uncertainty is indicated by vertical bars. For clarity, the systematic uncertainties are not shown, see Section~\ref{sec:systematic} for typical values.
 }
\end{figure*}

\begin{figure*}[tb]
\centering
 {\includegraphics[width=0.49\textwidth]{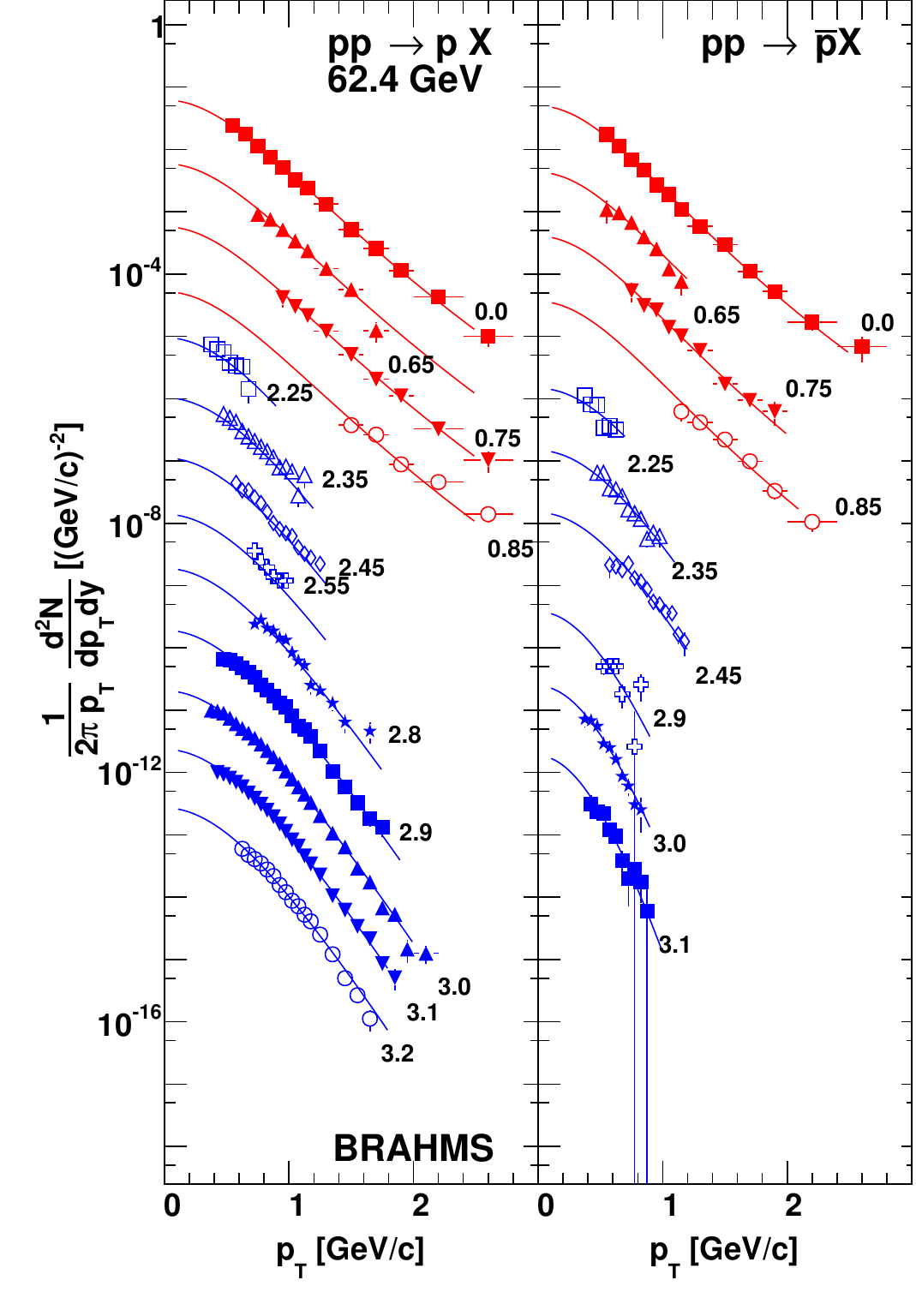}}
  {\includegraphics[width=0.49\textwidth]{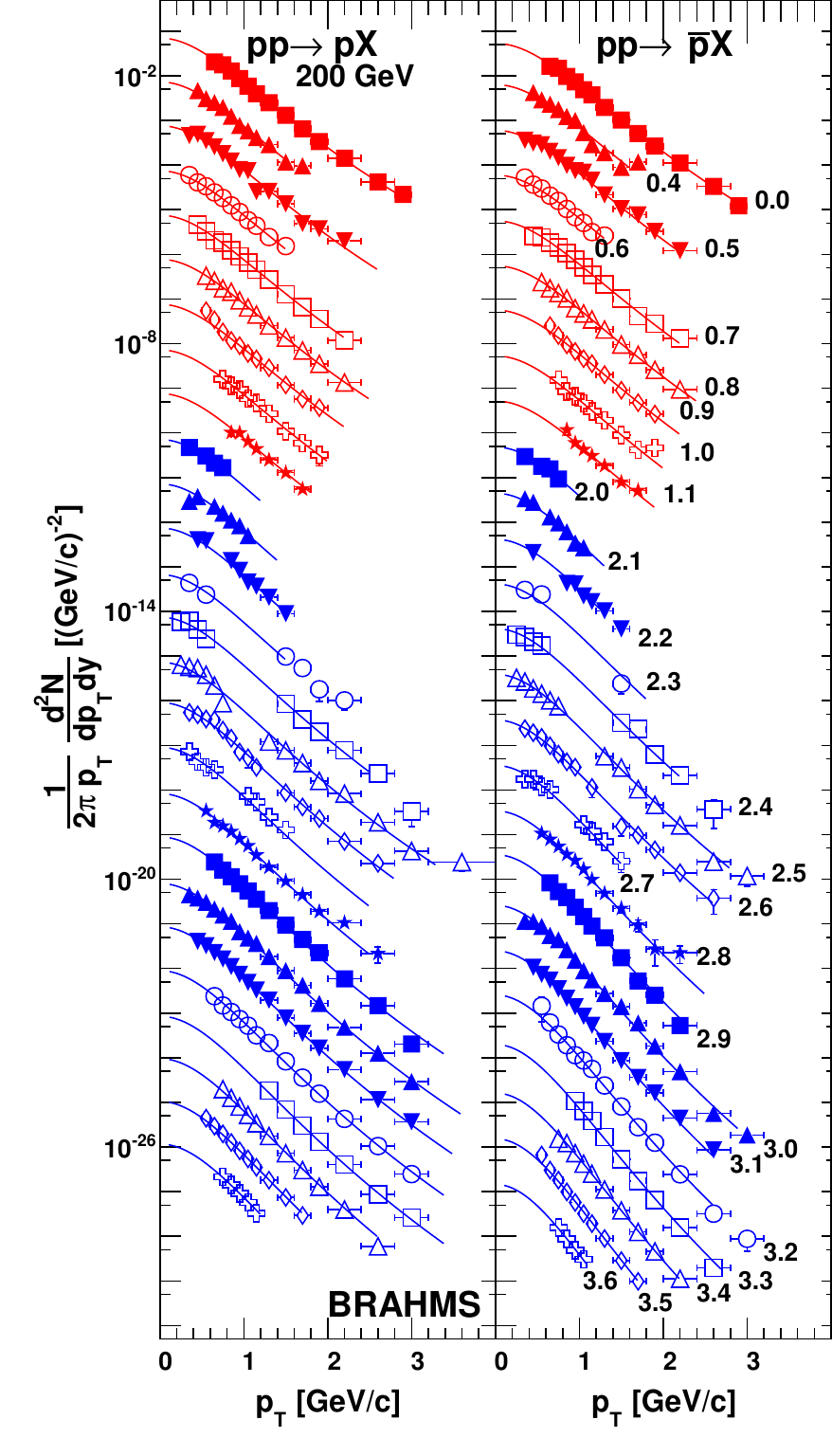}}
 \caption{\label{fig:protspectra} Invariant yields per event as a function of the transverse momentum \pt\ for protons and antiprotons  measured in \pp collisions at 62.4 and 200 GeV (two leftmost and two right most panels, respectively). 
 For each energy protons are shown to the left, and antiprotons to the right.
Rapidities are indicated at the end of each spectrum in \pp collisions at 62.4 and 200 GeV.
 Each rapidity bin is scaled down by a factor of 10 from the
 previous one. 
 The spectra for rapidities below 1.4 shown on top were measured with the MRS spectrometer (red), the rest were measured with the FS spectrometer (blue). The curves are the fits used to extract the integrated yields, see text for details.
 The statistical uncertainty is indicated by vertical bars. For clarity, the systematic uncertainties are not shown, see Section~\ref{sec:systematic} for typical values.
 }
\end{figure*}

\subsection{Transverse Momentum Spectra}

The invariant yields for pions produced in 62.4 GeV \pp collisions are shown in the left two panels of 
Fig.~\ref{fig:pionspectra}, the corresponding distributions for kaons are shown in Fig.~\ref{fig:kaonspectra}, and the ones 
for protons and antiprotons are shown in Fig.~\ref{fig:protspectra}.
The invariant cross sections are normalized to the total inelastic cross section of 36\,mb.
For clarity, the spectra extracted in different rapidity bins are scaled down by a factor of 10 starting
at mid-rapidity. 
The average rapidity value in each bin is listed to the right of each spectrum.
The solid curves in the figures are the results of the fitting with a Levy (Tsallis) function as discussed later in section \ref{sec:densities}.

The invariant spectra extracted from 200 GeV \pp collisions are shown in the right two panels of Fig.~\ref{fig:pionspectra} for pions, Fig.~\ref{fig:kaonspectra} for kaons, and Fig.~\ref{fig:protspectra} for 
protons and antiprotons.
The spectra that are derived from data that requires a CC vertex (MRS and FS-TOF) and thus represents a measurement of the NSD cross section, are normalized to the total inelastic cross section yield using the correction factors described earlier. Spectra in FS obtained requiring RICH do not need this normalization.
Each spectrum is scaled down by a factor of 10 with increasing rapidity bin. 

The spectra presented here have not been corrected for feed-down from the weak decays of $\rm{K}^0_s$, $\Lambda$ and higher mass hyperons. 
The main reason for not performing this correction for the proton and antiproton spectra is that the
$\Lambda$ and $\bar{\Lambda}$ yields and their dependencies on \pt and rapidity are not known, apart from at mid-rapidity in 200 GeV collisions where 
there are measurements by STAR~\cite{Abelev:2006cs}.

In general, the protons from weak decays are found at a lower average \pt\ than the parent $\Lambda$; the average \pt is approximately lowered by a factor $M_p/M_{\Lambda}$. 
Since most weak decays take place before reaching the first tracking detector (either TPM1 or T1) and we apply a fairly wide vertex constraint, most of the protons from the weak decays will be registered as indistinguishable from primary protons.
To quantify this further, we have performed detailed simulations of the $\Lambda$ spectrum with a spectral exponential form in $m_T = \sqrt{\pt^2 +m^2}$ and several inverse slope parameter values.
These simulated data are reconstructed in the complete BRAHMS analysis chain. 
The resulting proton spectra are integrated to extract the rapidity density yields at different rapidities.
From this analysis, it is determined that BRAHMS spectrometers measure ${\approx}90\%$ of the decay protons in the MRS, and ${\approx}80\%$ in the FS.
Hence, when comparing \dndy\ yields to models, a contamination from decay protons equal to the lambda yield multiplied by a rapidity-dependent
factor (0.90 or 0.80) times the relevant branching ratio (64$\%$) should be taken into account.
Thus, the integral of the measured yields closely matches the sum of direct proton spectra and those of decay protons from hyperons.
These feed-down corrections were also studied for BRAHMS in detail for Au+Au reactions at 200 GeV~\cite{BRAHMS:auau2005PRC,Bearden:2003hx}.
To compare the \pt dependence of spectra, one can add the model proton spectra with a derived decay proton spectrum from lambdas where the \pt is scaled by the mass ratio using the factors given above.
The same procedure applies to $\bar{\Lambda}$ and antiproton spectra and yields.
This procedure is used for the \pythia and \epos comparisons discussed later in this paper.

\subsubsection{Comparison with ISR data}

\begin{figure*}[htb]
\centering
{\includegraphics[width=0.49\textwidth]{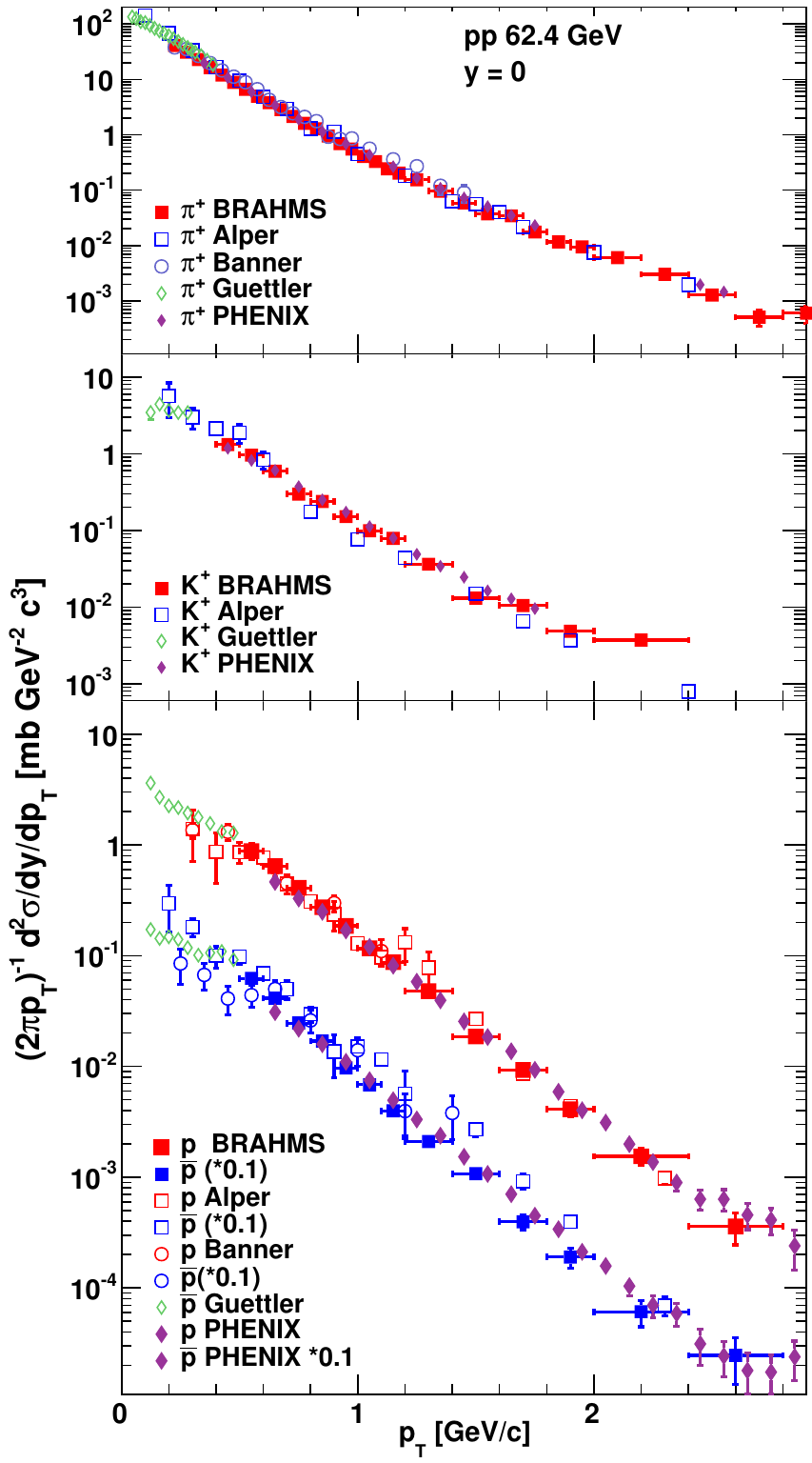}}
{\includegraphics[width=0.49\textwidth]{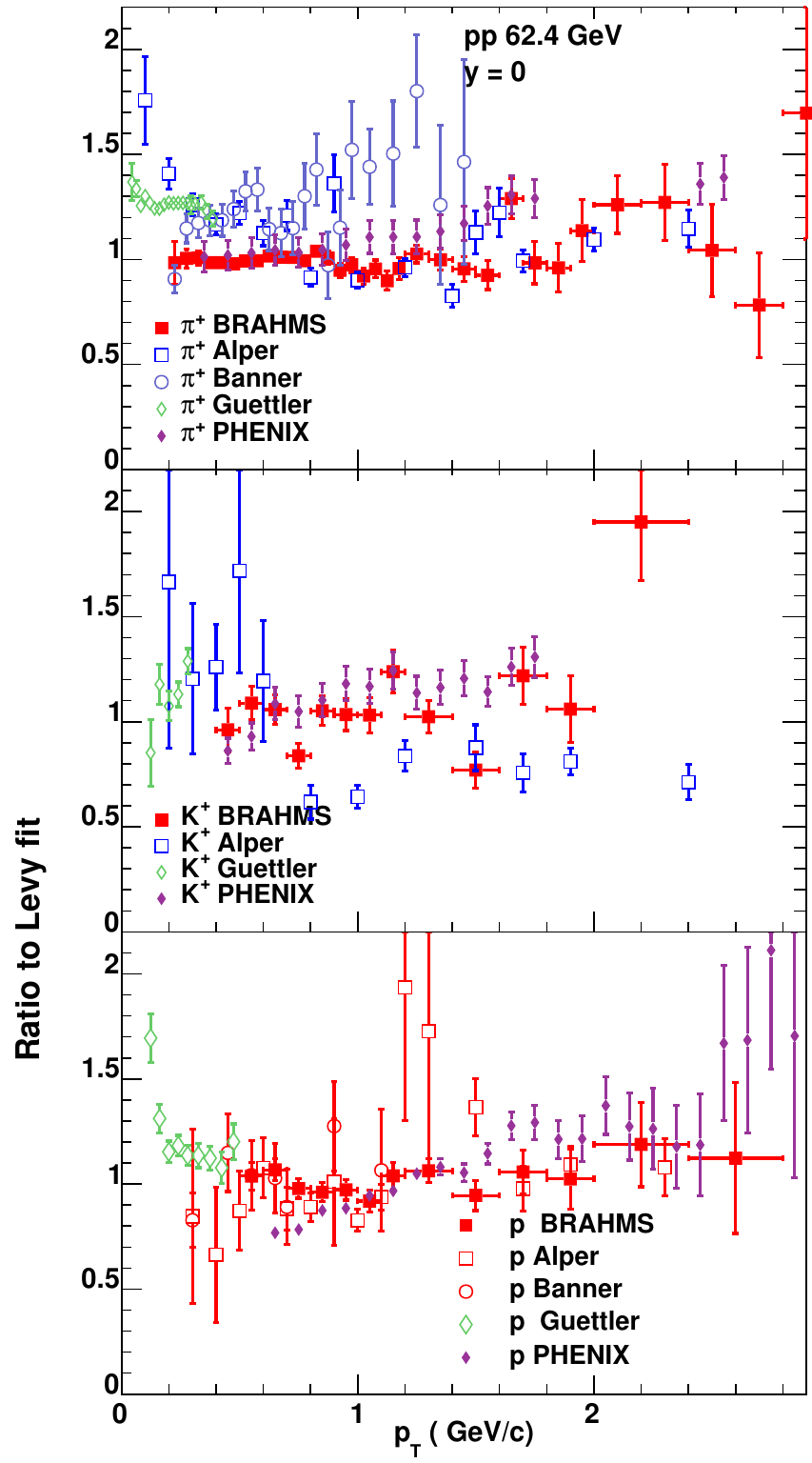}
} \caption{\label{fig:isrcomp} 
Comparison of ISR and PHENIX data with present 62.4 GeV data at mid-rapidity. The left panels is a comparison with the spectra, and the right panels show a comparison with the Levy fit of the BRAHMS data. 
In each of these panels the top-sub panel is for \piplus, the middle is for \kplus, and the bottom sub-panel for protons and \pbar. The ratios for \pbar have been omitted for clarity.
The \pbar cross sections in the left panel have been divided by a factor of 10 for clarity. The PHENIX data are the non feed-down corrected data.}
 
\end{figure*}

Proton+proton collisions were studied extensively at the CERN ISR collider at energies from \sqrts = 29 to 62.4 GeV. 
Of particular relevance for this work are the results at 62.4 GeV from several experiments. 
The BRAHMS dataset has in general better statistics than the previous ISR measurements.
At mid-rapidity, differential cross sections from BRAHMS are compared with data from Alper~\cite{Alper:1975jm}, Guettler~\cite{Guettler:1976fc}, and Banner~\cite{Banner:1977pe}. 
Overall there is good agreement between the pion, kaon, proton and antiproton measurements as shown in the upper, middle and lower sub-panels of the left panel Fig.~\ref{fig:isrcomp}.

The cross sections from Guettler extend to lower values in \pt and show a turnover consistent with a thermal description of the spectra. 
This demonstrates that the assumption of a power-law spectrum 
 is not justified, and that a description using, e.g., the Levy function \cite{Wilk:Levy,STAR:Levy,Adare:2010fe} as discussed in the next section, which takes this low-\pt feature into account, is more appropriate.

The lower left panel of Fig.~\ref{fig:isrcomp} compares BRAHMS proton and antiproton cross sections with those of several ISR 
experiments~\cite{Alper:1974xp,Guettler:1976ce,Banner:1977pe}. 
The right panels show the ratio of the spectra from the different experiments to the Levy function fitted to the BRAHMS data.
The agreement for protons is quite good, but all the ISR \pbar~data are clearly above the BRAHMS results. 
Some of the ISR \pbar/p ratios even have values above 1, which is not expected.

Comparisons are also shown for mid-rapidity spectra to the published PHENIX data from \pp collisions at 62.4 GeV~\cite{Adare:2011vy} shown as the dark purple diamonds in Fig.~\ref{fig:isrcomp}. 
Note that the PHENIX data for protons and antiprotons shown are not feed-down corrected.  Overall, the agreement at 62.4 GeV is very good within the systematic uncertainties quoted by each experiment. 

Fewer data from the ISR are available at forward rapidity from the 62.4 GeV \pp collisions. 
Negative pion cross sections as a function of \xF were published vs.\ energy by Albrow~\cite{Albrow:1973} at energies up to 52.4 GeV at angles relative to beam rapidity $\tan(\theta)= 2.66 /\sqrt{s}$ that fall on a common curve.
 The BRAHMS data have a region of overlap with  these data at rapidity 3.4 and \xF $\sim 0.3$ and we have checked (not shown) that the two data sets agree well within the systematic uncertainties of ${\approx}10\%$.

\subsubsection{Comparison with 200 GeV \pp RHIC data}

\begin{figure*}[htb]
\centering

{\includegraphics[width=0.49\textwidth]{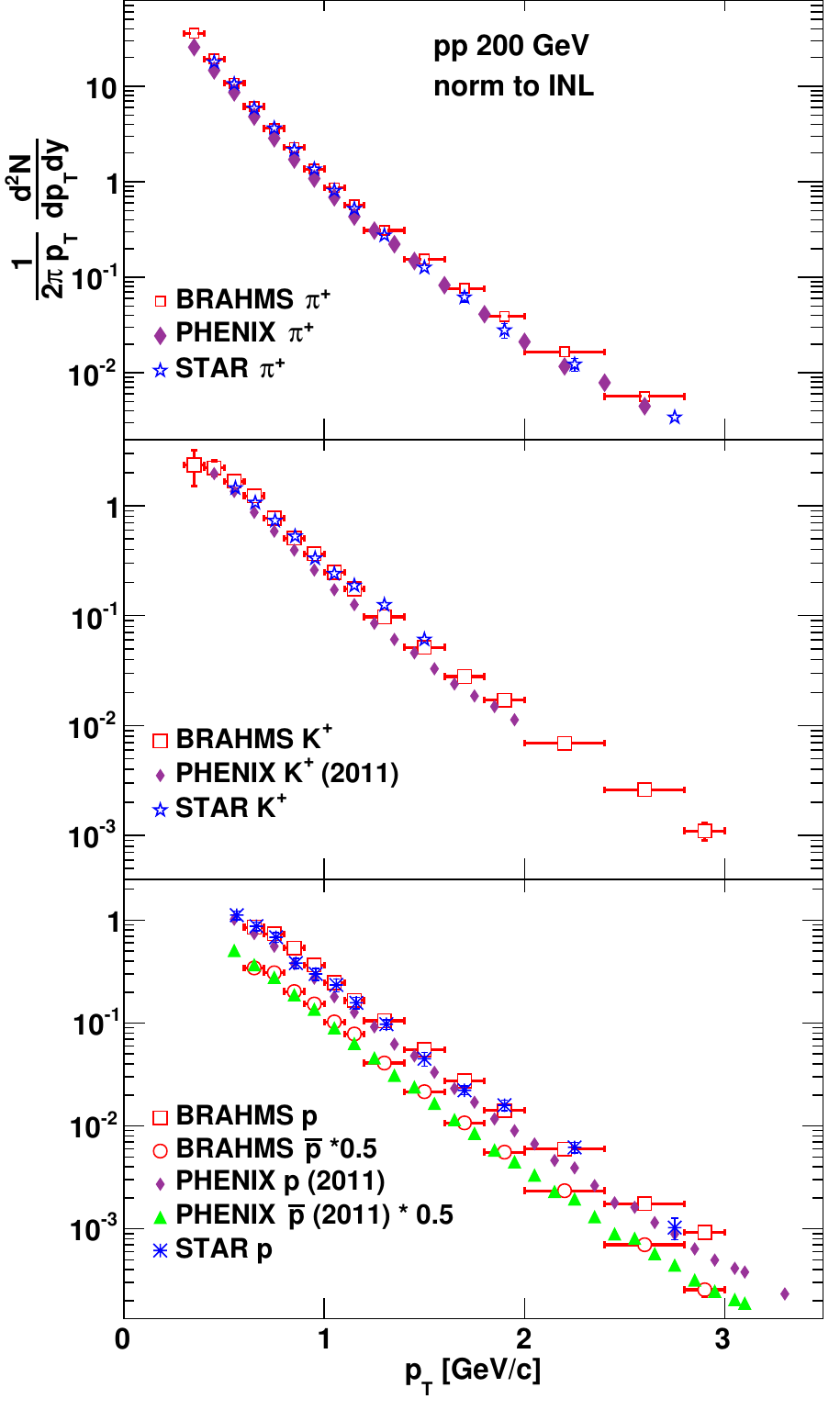}}{\includegraphics[width=0.49\textwidth]{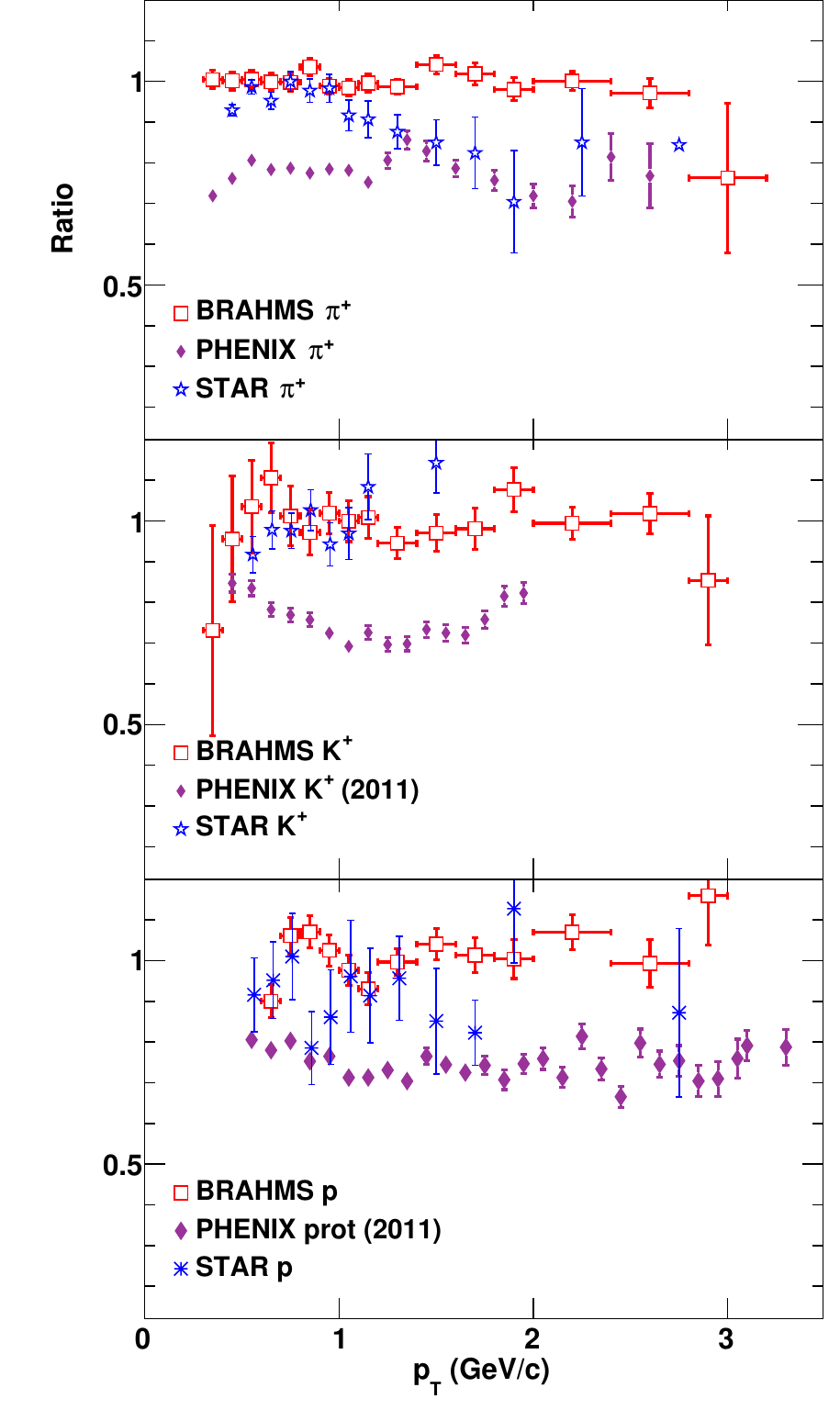}}
 \caption{\label{fig:rhiccomp}Comparison of STAR and PHENIX data with present 200 GeV data at mid-rapidity. The left panels is a comparison with the spectra, and the right panels show a comparison with the Levy fit of the BRAHMS data. The STAR data has been re-normalized to represent the inelastic cross sections. In each of these panels the top sub-panel is for \piplus, the middle is for \kplus, and the bottom sub-panel for protons and \pbar. The ratios for \pbar have been omitted for clarity. The \pbar cross sections have been divided by a factor of 2 for clarity.
 The PHENIX data are the non feed-down corrected.
}
 
\end{figure*}
 
Both STAR and PHENIX have published results for mid-rapidity identified charged particles in \pp collisions at 200 GeV \cite{STAR:2008ez, Adare:2011vy}. The PHENIX and BRAHMS data are both normalized to the inelastic cross section cross section of 42 mb. The published STAR data are normalized to NSD, but have been renormalized to the inelastic  cross section for this comparison.
The BRAHMS spectra at mid-rapidity cover the \pt-range from 0.2 to 2.0-3.0\,\gevc\ and within that range the spectra, from all three RHIC experiments, have similar shapes as shown in  Fig.~\ref{fig:rhiccomp}.
Considering the fact that the STAR pion data are corrected for weak decays, the BRAHMS pion distributions appear to be about 10\% higher than STAR in the low \pt-region.

Comparing the pion and kaon spectra we find that BRAHMS data are about 25\% higher than PHENIX, but with a similar \pt-dependence. 
The systematic uncertainty on the respective Vernier scanned cross sections is 8\% for PHENIX and 13\% for BRAHMS.
There is  a slight tension between these results at a $1.5\sigma$ level.
An additional cross check has been performed by comparing the measured charged pseudo rapidity density, $dN/d\eta$, from UA5~\cite{Alner:1986_AU5} and PHOBOS~\cite{Alver:2010ck} with the \dndy  for pions as measured by PHENIX and BRAHMS. 
By studying the results of \pythia calculations using several different tunes it is observed that the ratio $(dN/dy(\pi^+))$/$(dN/d\eta)$ is ${\approx}0.5$ at y=0.
The inelastic charged particle yield $dN/d\eta$ at y=0 is $2.2\pm 0.2$ for UA5, and  $2.25 + 0.37 - 0.30$ from PHOBOS, thus the predicted pion multiplicity should be $1.1\pm 0.1$ ($1.13+0.18-0.15$).
BRAHMS has measured a value for \dndy of $1.25\pm0.18$ while the PHENIX value is 0.82-0.92 depending on the extrapolation used.
So the pion yield reported here agrees with the expectation for the charged particle $dN/d\eta$ within systematic uncertainties, albeit slightly higher, whereas the PHENIX results are $20$--$25\%$ lower. If we make a similar comparison of $dN/d\eta$ with the STAR pion data converted to inelastic cross section, the STAR result agrees within systematic uncertainties.

\subsubsection{\label{sec:particleratio}Particle Ratios}
\begin{figure*}[tb]
\centering
 {\includegraphics[width=\linewidth]{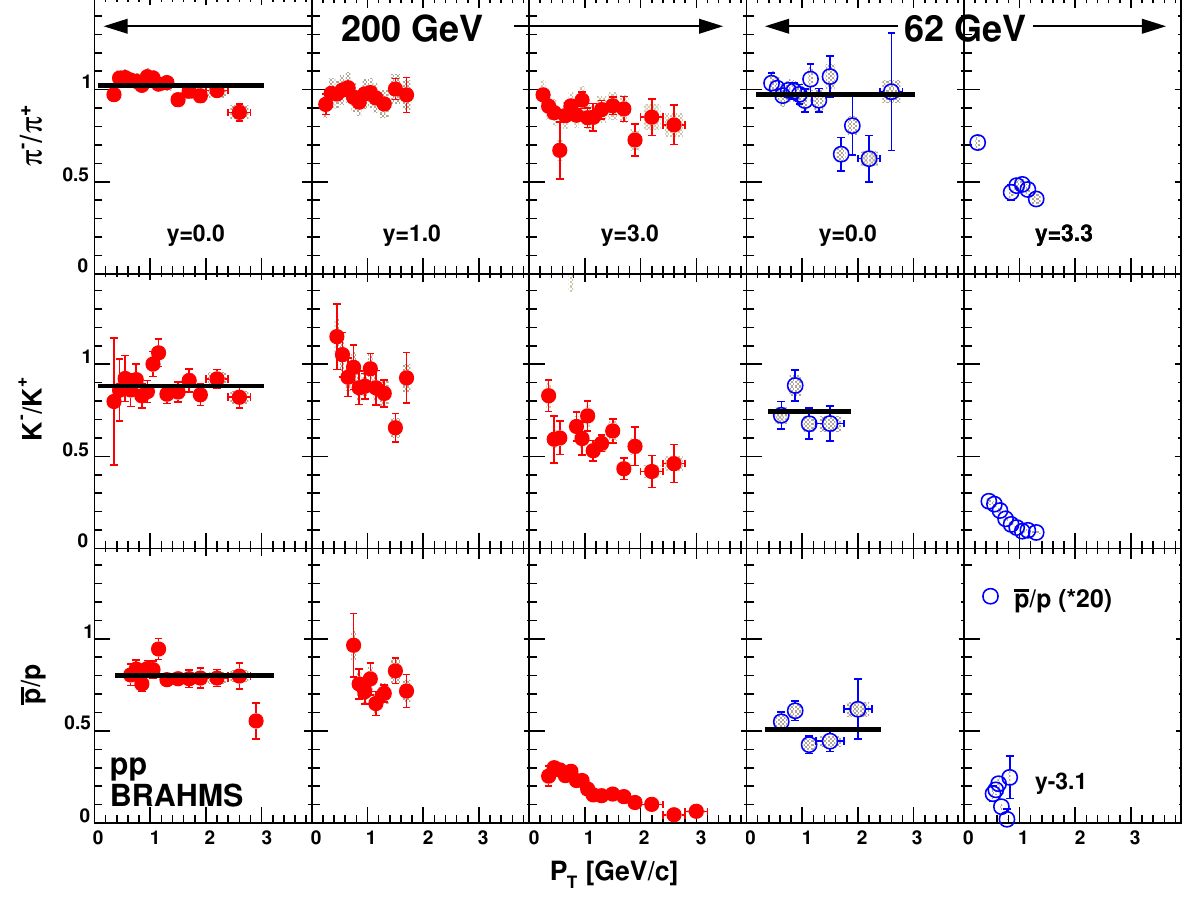}}
 \caption{\label{fig:partRatios200_62} 
 Negative-to-positive particle ratios for pions, kaons, protons at 
200 GeV (left three columns) and 62.4 GeV (right two columns).  The shaded bars, which are hardly visible, represent the systematic uncertainties.
 }
\end{figure*}

The ratios of like-particle spectra measured in 200 GeV \pp collisions, at selected rapidities, are shown
in the three left columns of Fig.~\ref{fig:partRatios200_62}. 
The ratios at mid-rapidity are computed by taking the ratios of spectra from one magnetic field polarity to spectra measured at the opposite magnetic field polarity in order to reduce the systematic uncertainties due to detector response.
At mid-rapidity there is no \pt dependence in the measured range. 
The \pt averaged values are $\pi^{-}/\pi^{+} = 1.02 \pm 0.01 \text{(stat)} \pm 0.04 \text{(syst)}$, 
$K^{-}/K^{+} = 0.882 \pm 0.02 \text{(stat)} \pm 0.04 \text{(syst)}$, and
$\bar{p}/p = 0.800 \pm 0.013 \text{(stat)} \pm 0.04 \text{(syst)}$, in good agreement with previous results \cite{Bearden:2004ya,Adler:2006xd,Abelev:2006jr}
At high rapidity the like-particle ratios are lower and show a strong decreasing \pt dependence.

The like-particle ratios for the 62.4 GeV data are shown in the two right-most columns of Fig.~\ref{fig:partRatios200_62}. 
As is the case for the 200 GeV data, the pion ratios have values close to 1 at mid-rapidity and decrease rapidly at forward rapidities. 
The kaons show a similar behavior. 
Within the statistics of our measurements we do not observe a \pt~dependence of the ratios up to \pt of 2\,\gevc\ at mid-rapidity.
The \pt~averaged values are $\pi^{-}/\pi^{+} = 0.974 \pm 0.015 \text{(stat)} \pm 0.04 \text{(syst)}$, 
$K^{-}/K^{+} = 0.742 \pm 0.042 \text{(stat)} \pm 0.04 \text{(syst)}$, and
$\bar{p}/p= 0.51 \pm 0.03 \text{(stat)} \pm 0.04 \text{(syst)}$.
At $y\sim3$ a strong \pt-dependence is observed and with very small $\bar{p}/p$ ratios.

\begin{figure}[tb]
\centering
 {\includegraphics[width=3.2in]{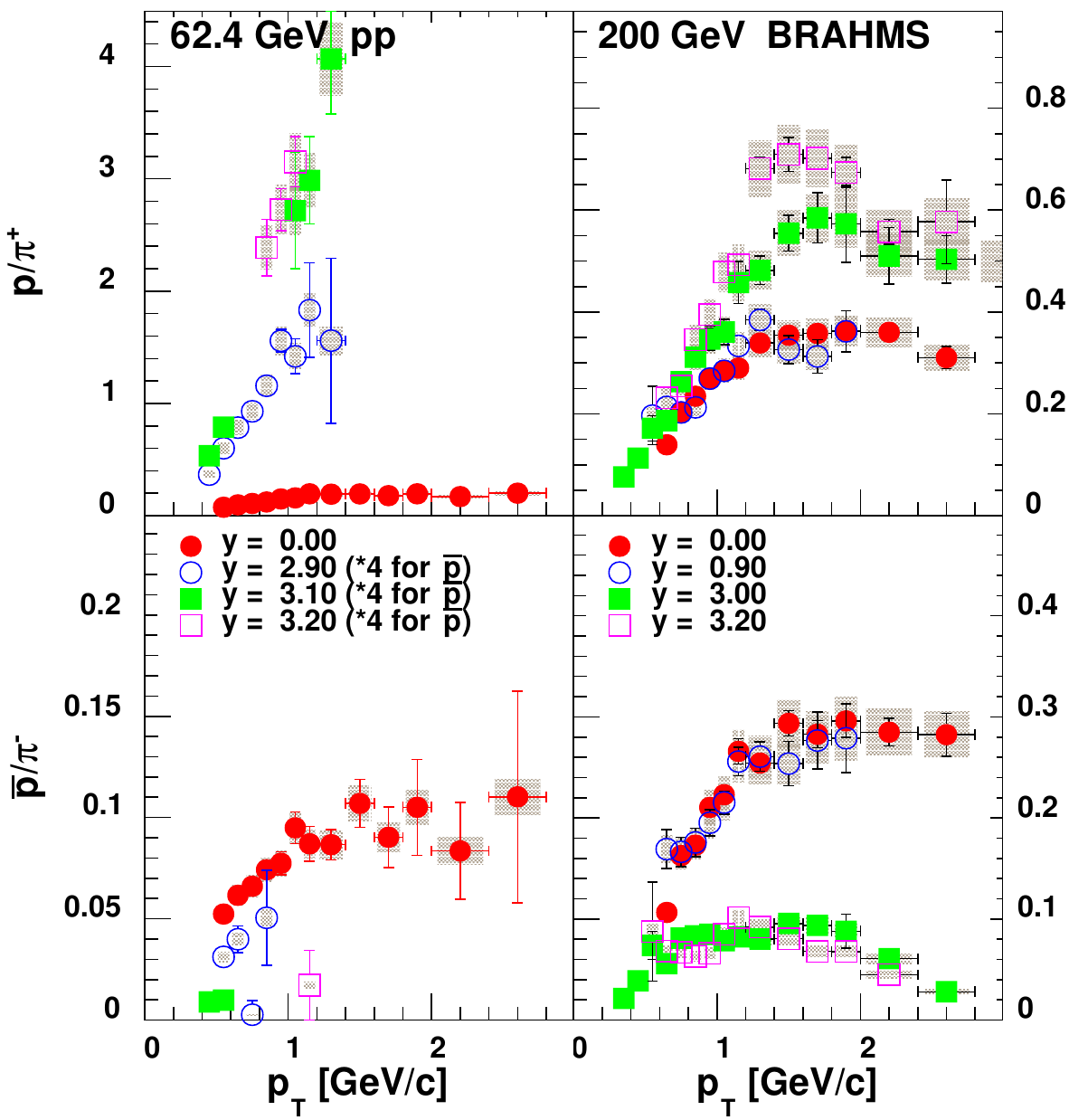}}
 \caption{\label{fig:PoverPi} 
 Particle ratio of protons over pions for $p/\pi^{+}$ (upper
 panels) and $\bar{p}/\pi^{-}$ (lower panels) for rapidities
 indicated by the legend. Data from 200 GeV are in the right panels,
 the 62.4 GeV data in the left.  The shaded bars represent the systematic uncertainties.
 }
\end{figure}

Figure~\ref{fig:PoverPi} shows the ratios of $p/\pi^+$ and 
$\bar{p}/\pi^-$ for the data at 62.4 and 200 GeV for a few selected rapidity bins. 
The behavior of the 200 GeV ratios as a function of \pT does not change much for small shifts in rapidity at $y\sim 0$ or $y\sim3$, but
shows strong variations when going from $y \sim 0$ to $y \sim 3$ especially for the $\bar{p}/\pi^-$ ratio.
The left panels of Fig.~\ref{fig:PoverPi} show the ratios extracted from the 62.4 GeV measurements. 
The increase of the $p/\piplus$  ratios as a function of rapidity is much stronger than the one found at 200 GeV whereas the $\bar{p}/\pi^{-}$ are much smaller.
A more detailed discussion can be found in~\cite{Arsene:2009nr}.

\begin{figure}[tb]
\centering
 {\includegraphics[width=3.2in]{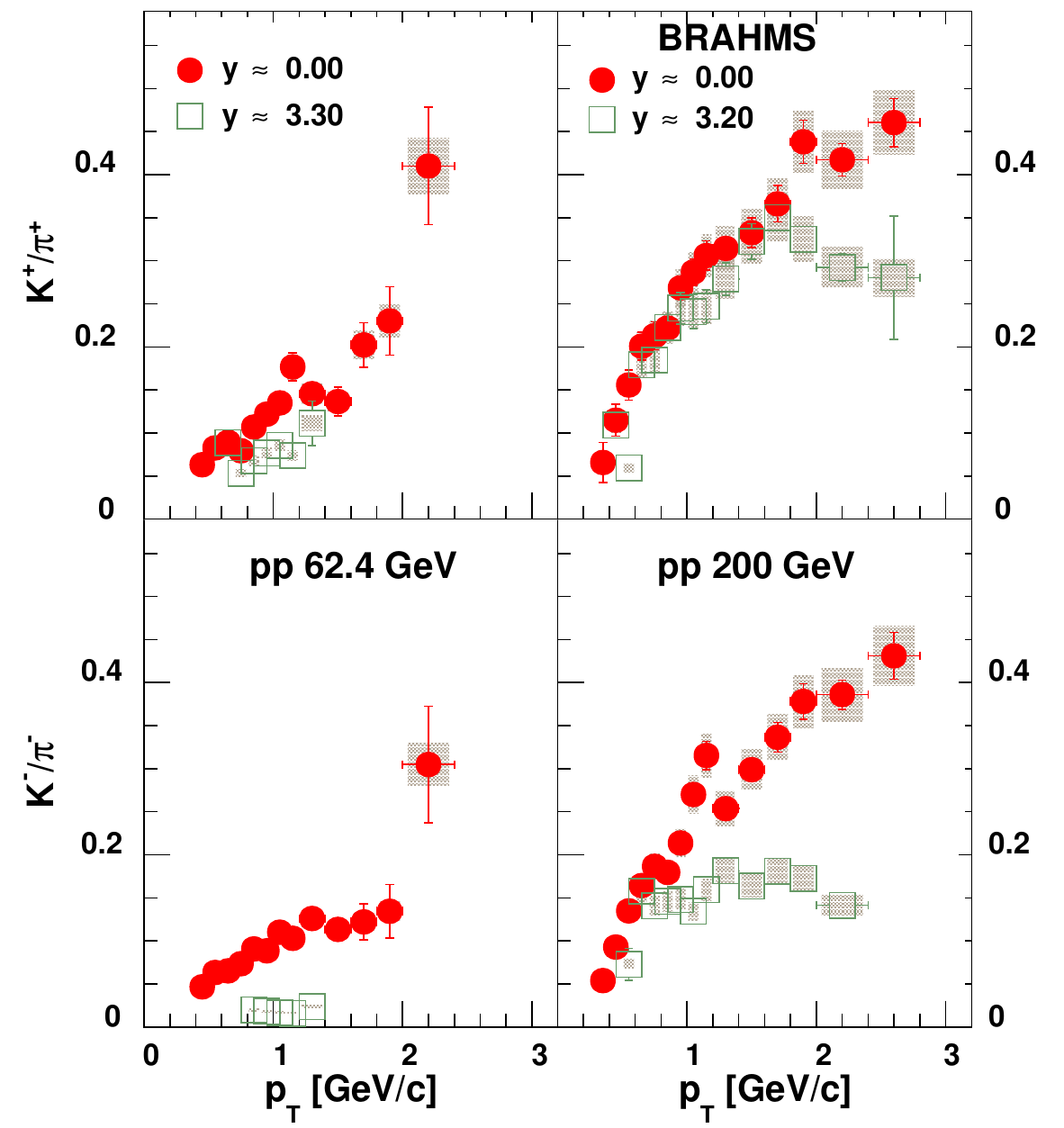}}
 \caption{\label{fig:KoverPi} 
 Kaon-over-pion ratios for $\rm{K}^{+}/\pi^{+}$ (upper
 panels) and $\rm{K}^{-}/\pi^{-}$ (lower panels) for rapidities
 indicated by the legend. Data from 200 GeV are in the right panels,
 the 62.4 GeV data in the left panels. The shaded bars represent the systematic uncertainties.
  }
\end{figure}

Figure~\ref{fig:KoverPi} shows the $\kplus/\piplus$ ratios (top panels) and
the $\kminus/\piminus$ ratios (bottom panels) for the 62.4 (left panels) and 
200\,GeV collisions (right panels) at mid-rapidity and at forward rapidities.
A much stronger decrease with rapidity has been measured in the $\kminus/\piminus$ ratios than the $\kplus/\piplus$ ratios, most markedly at the lower energy.

\subsection{\label{sec:densities}Rapidity Distributions}

Because the measured \pt-spectra do not cover the entire \pt\ range, extrapolations towards lower and higher \pt\ values are necessary to extract the rapidity densities. 
In particular, the results obtained for pions depend strongly on the extrapolation towards zero \pt\ due to their low average \pt\ of about 300-400\,\mevc, while BRAHMS coverage 
typically extends only
down to ${\approx}350$ \mevc over most of the covered rapidity range (200 \mevc only for few of the rapidity bins).  
Different functional forms have been proposed to perform such extrapolations in \pt. 
The Levy function \cite{Wilk:Levy,Tang:2008ud,STAR:Levy,Adare:2010fe} is now often used for relativistic heavy-ion reactions, since
it combines the feature of a power-law behavior at high-\pt\ with that of an exponential in $m_T$ at low \pt.
The functional form used to analyze the invariant \pt-distributions is given by Eq.~\ref{eq3}.
\begin{eqnarray}
\frac{1}{2\pi \pt}\frac{d^2N}{dyd\pt}&=&\frac{1}{2\pi}\frac{dN}{dy}\frac{(n-1)(n-2)}{nT(nT + m_0(n-2))}\nonumber \\
&\times&\Big(1+\frac{(m_T-m_0)}{nT}\Big)^{-n}, \label{eq3}
\end{eqnarray}
where the fit parameters are $\dndy$, $n$, and $T$.
To minimize the effect of having a different \pt coverage at different rapidities, a global fit to spectra at all rapidities is performed to extract the yields. 
The mean \pt and the high \pt tail are known to decrease with increasing rapidity, and we are using this as a guidance for a global fit, that results in additional measurement points.
The procedure was checked by comparing results for \dndy and \meanpt with results from individual fits that have good \pt coverage.  
The Levy function parameters $n$ and $ T $ are assumed to be slowly varying functions of rapidity, i.e., $T = T_0 +  ay^2$ and 
$n = n_0 + by^2$.
In this case the free parameters are $T_0, n_0, a, \rm{and} $
$b$. 

Such a fit reduces the systematic variations that would otherwise be present in the parameters  as a consequence of the different \pt coverage at the different rapidity settings. 
The drawback of this procedure is a slightly higher $\chi ^{2}$ for each setting resulting from the additional constraint on the rapidity dependence.
These values are still satisfactory, with $\chi^2/\text{NDF}$ of 1--2 for each rapidity setting. 
The fits to the 200\,GeV invariant yields are presented in Figs.~\ref{fig:pionspectra}, \ref{fig:kaonspectra}, and \ref{fig:protspectra}. 

The systematic uncertainty from the extrapolation to zero \pt is estimated by varying the fitting ranges and by assuming different function form for the spectral shape e.g. pure exponential in \pt and in \mt. 
In general the $\chi^{2}/ NDF$ for alternate shapes were worse and give less weigh for the estimates.
Overall systematic due to this effect used was $12\%$ for pions, and $8\%$ for kaons and protons due to the larger \meanpt.

At 62.4\,GeV, a similar global fit that includes spectra from  all rapidity bins at once is not possible, because the spectral shape at the most forward angles is influenced by the closeness to the kinematic limit and changes rapidly with rapidity.
Instead, the data set is divided into two groups: one around mid-rapidity and the other at high rapidity. Combined fits for settings within these groups are then performed with a procedure similar to the 200\,GeV data.
The 62.4\,GeV data have far fewer data points, and less coverage in rapidity and \pt due to the short running time at 62.4 GeV. 
Such limitations restrict the extraction of \dndy.
Kaons, in particular, have the smallest coverage in \pt for the measured rapidities.

\begin{figure}[tb]
 \centering
 {\includegraphics[width=3.2in]{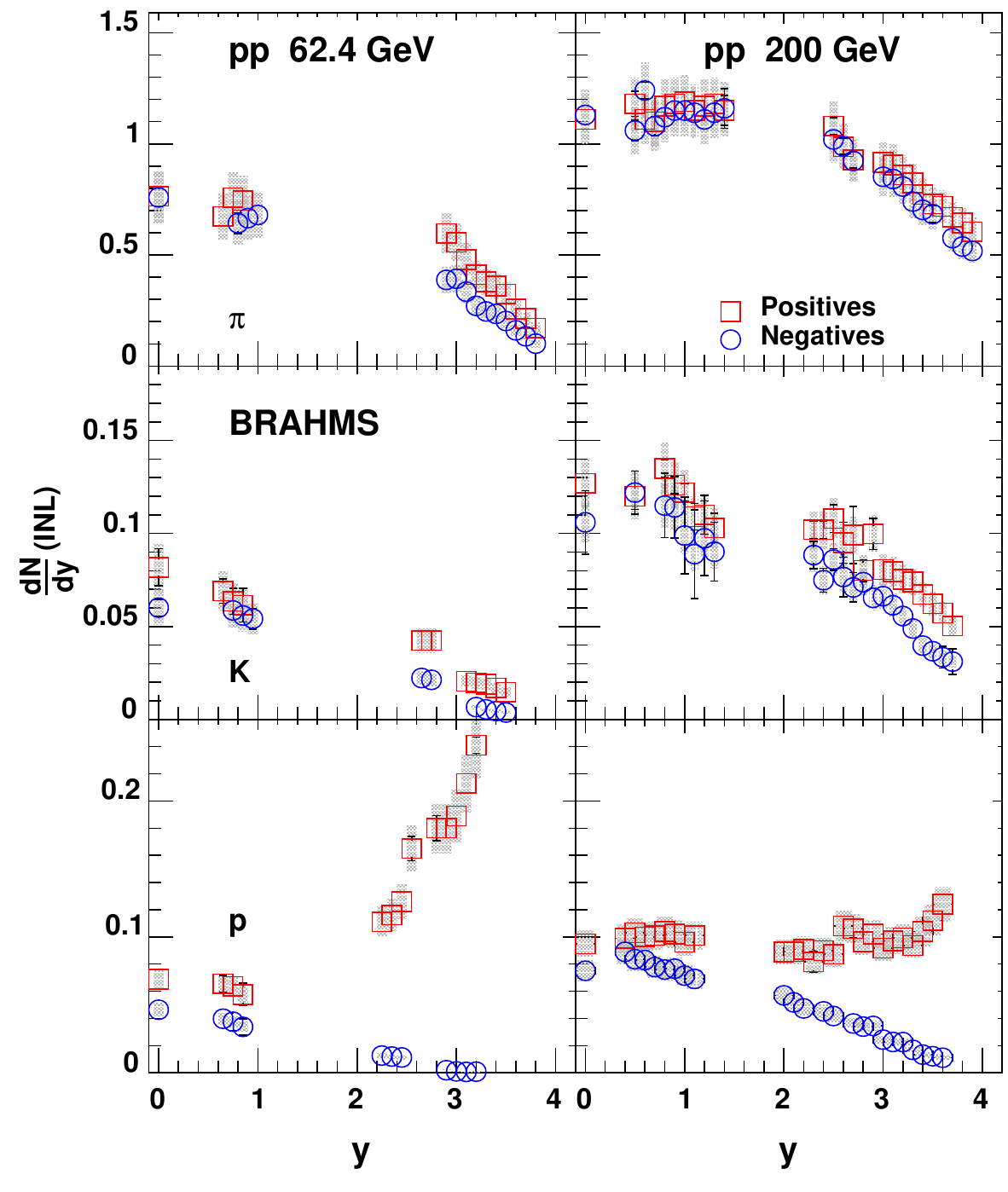}}
 \caption{\label{fig:dndy_200_62} 
  Invariant rapidity densities for pions, kaons and protons at 62.4 (left) and 200 GeV (right). 
  Open squares (red) represent the positive particles and open circles (blue) represent the negative particles.
  The shaded bars represent the systematic uncertainties, not including the overall normalization.
}
\end{figure}

\begin{figure}[tb]
\centering
 {\includegraphics[width=3.2in]{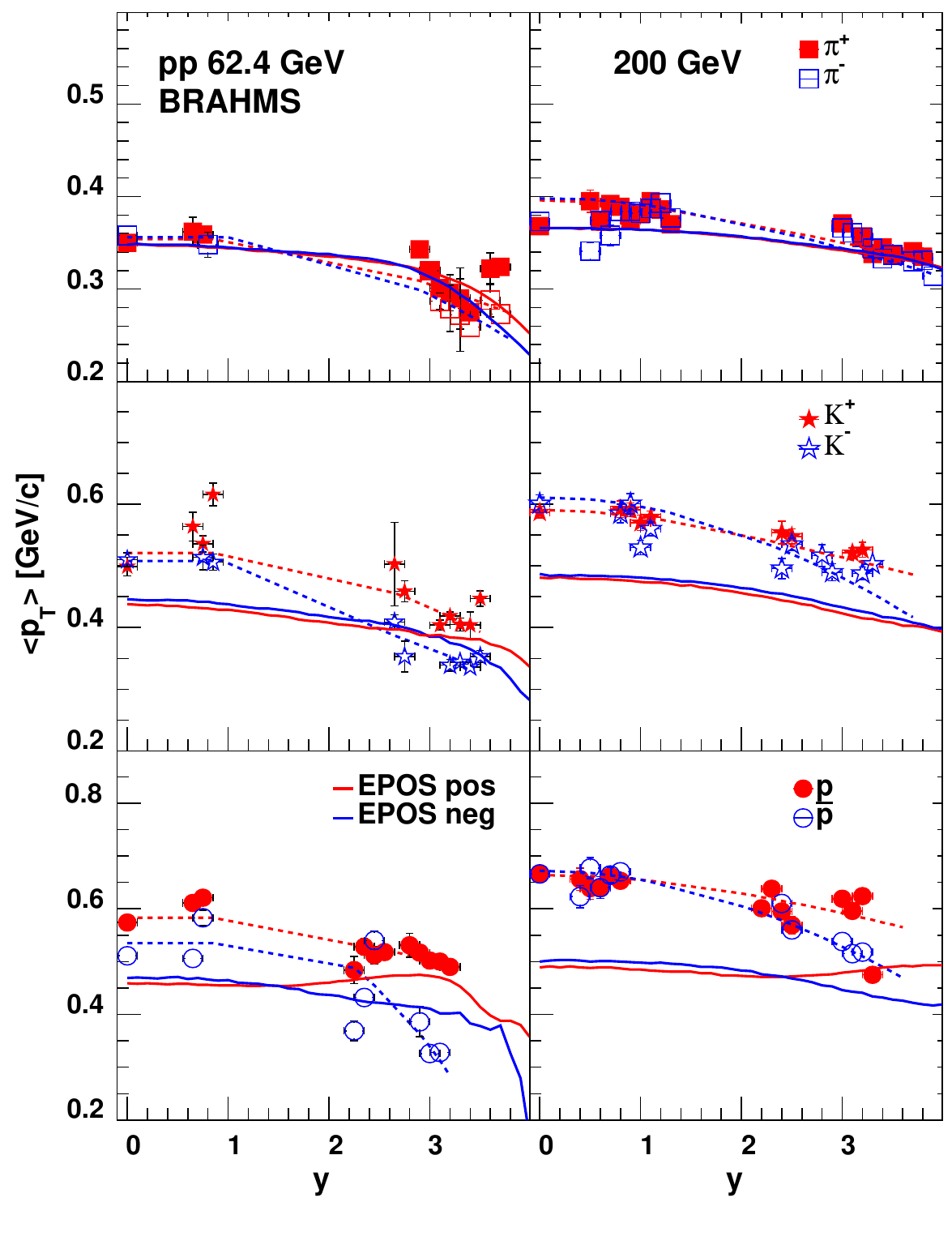}}
 \caption{\label{fig:meanptall} 
 \meanpt\ for pions, kaons, protons and antiprotons from 62.4\,GeV (left) and 200\,GeV (right) vs.\ rapidity. The data points represent results obtained from Levy fits to individual spectra with good \pt coverage. The dotted lines are \meanpt as deduced from the global rapidity fits. The solid lines are the \epos calculations for positive (red) and negative (blue) discussed later in the text. Comparisons with \pythia are not shown for clarity reasons but are in general similar to the \epos predictions. }
\end{figure}

The extracted rapidity densities, \dndy from \pp collisions at 200\,GeV, are shown in the right panels of Fig.~\ref{fig:dndy_200_62}. 
Results for pions, kaons and protons are shown in the top, middle, and bottom panels, respectively. 
Positive particles are presented with open red squares and negatively charged particles are shown with open blue circles. 
An almost flat distribution extending up to $y\sim1$ followed by a decrease at higher rapidities is found for both pions, kaons, and antiprotons. 
Protons deviate from that trend showing an increase above $y\sim2$ at both energies.

The left panels of Fig.~\ref{fig:dndy_200_62} show the 62.4\,GeV rapidity densities. 
The 62.4\,GeV distributions are in general lower and narrower than the 200\,GeV data as discussed later in section \ref{sec:mult}. 
 The pions (at 62.4\,GeV) show a similar behavior as the 200\,GeV data, namely a flat distribution up to $y\sim1$ and then a decrease. 
The protons, in contrast, show a much stronger increase at the forward rapidities than the 200\,GeV data. 
This behavior is a consequence of the increased proximity to the beam rapidity at the lower energy. 
The antiproton \dndy distributions show a much stronger decrease at high rapidity than for the other particles.
Using the parameters extracted from the individual Levy fits, the average transverse momentum, \meanpt, has been derived for each of the particle species. 
The \meanpt vs.\ rapidity distributions are  shown in Fig.~\ref{fig:meanptall} with the markers. 
The right panels show the data for $\sqrt{s}=200$ GeV, and the left panels show the data for the $\sqrt{s}=62.4$ GeV. 
The dashed lines represent the \meanpt values derived from the global fit parameters.
We note that these two quantities in general agree.

The magnitude of the average transverse momentum, \meanpt, 
shows an increasing trend from 62.4 to 200 GeV, mild for pions and more pronounced for heavier hadrons. 
At both energies the \meanpt for all species is approximately constant near mid-rapidity, and drops towards the forward rapidities.

\subsubsection{\label{sec:longscaling}Longitudinal Scaling}

It was conjectured by Benecke et al.\ in Ref.~\cite{1969PhRv..188.2159B} that particle production near beam rapidity should be independent of beam energy when cross sections are measured relative to the beam rapidity, i.e., vs.\ the variable $y-y_{beam}$. 
Equivalently, this is also a consequence of Feynman scaling \cite{ Feynmanscaling}. 
The original expectation was that such behavior should also be present for differential cross sections, i.e., ${\frac{d^2N}{{d\pt}dy}} (y-y_{beam},\pt)$.
In most of the studies related to this conjecture, the dependence of the rapidity density \dndy or pseudo-rapidity $dN/d\eta$  have been explored.

This phenomenon, called  ``longitudinal scaling'', was  demonstrated in  a survey of ISR data for an energy range from 26 to 52 GeV in the review papers of Giacomelli and Jacobs \cite{Giacomelli:1979nu,Allaby:1974rx}.
This longitudinal scaling has also been observed in cosmic-ray data and in AA collisions (see for example \cite{1978NuPhB.142..445O} and \cite{busza:2011pc} and references therein). 
In this section, such scaling is explored with the BRAHMS \pp data at 62.4 and 200\,GeV.

\begin{figure}[tb]
\centering
 {\includegraphics[width=3.2in]{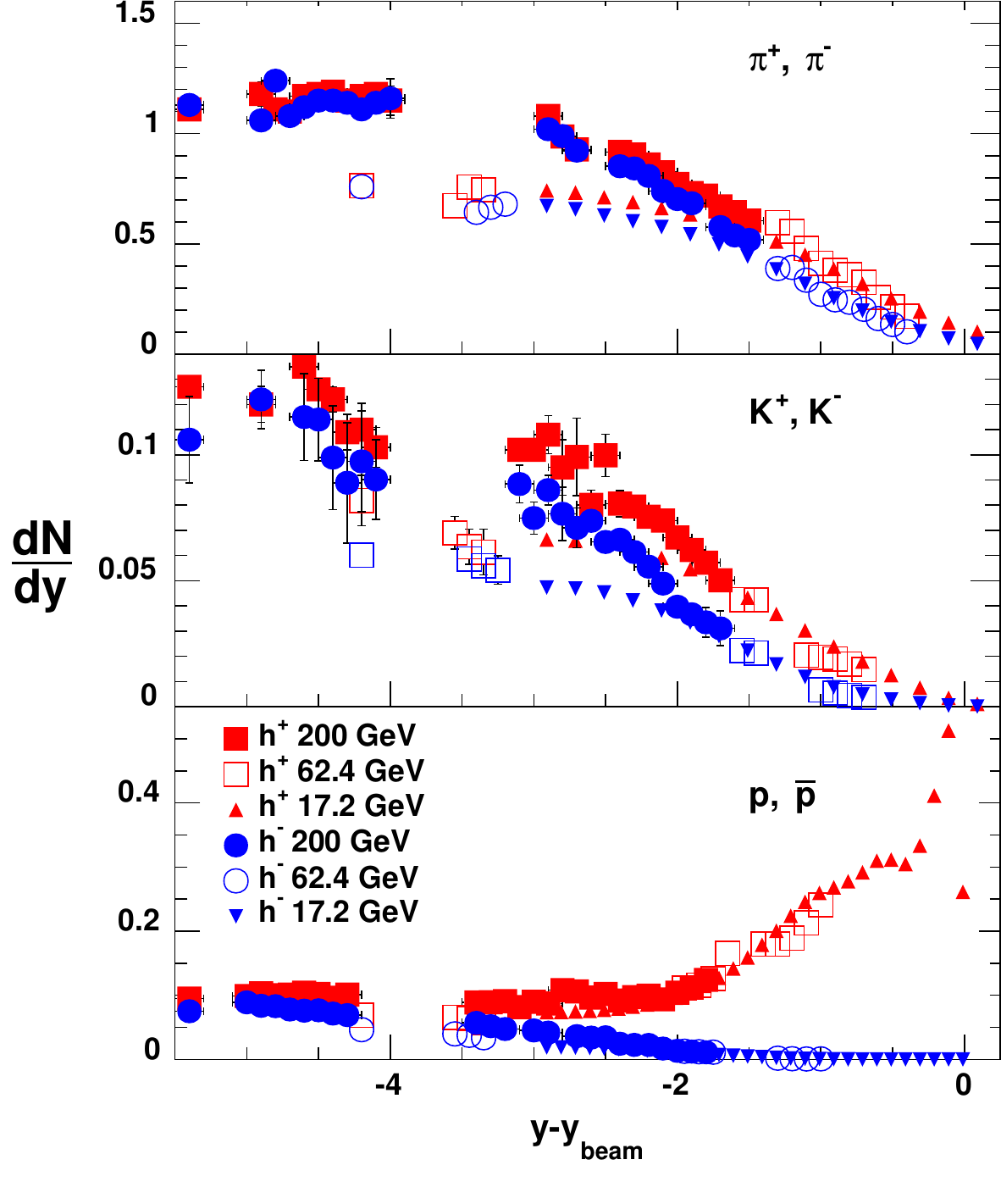}}
 \caption{\label{fig:partscaling} 
 Rapidity densities for $\pi^+$, $\pi^-$, K$^+$, K$^-$, $p$ and $\bar{p}$ at 62.4 and 200\,GeV (BRAHMS) and 17.2 GeV (NA49) plotted vs.\ $y-y_{beam}$.
 The data illustrates the longitudinal scaling for \pt integrated yields. Uncertainties are statistical only on this figure.
 }
\end{figure} 

Figure \ref{fig:partscaling} shows the same rapidity density distributions as shown in
Fig.~\ref{fig:dndy_200_62}, but plotted vs.\ the difference in rapidity from the beam rapidity, $y - y_{beam}$.
The top panel shows the \dndy for pions, the middle panel shows the distributions for kaons, and the bottom panel shows the same quantities for $p$ and $\bar{p}$. 
This compilation includes BRAHMS data from 62.4 and 200\,GeV,  and pion, kaon, and proton data from NA49~\cite{Alt:2005zq,Anticic:2009wd} at $\sqrt{s}=17.2$ GeV. 
We note that NA61 also remeasured the 17.2 GeV data \cite{Aduszkiewicz:2017sei} and these data agree well with the NA49 data. 
It is worth noting that for all energies the data tend to overlap near the beam rapidity of their respective system.
This scaling suggests that particle production near the beam rapidity is indeed governed mostly by the variable $y - y_{beam}$
regardless of the energy of the system, as conjectured. 
This feature appears to be true for the range $17.2 \leq \sqrt{s} \leq 200$\,GeV,  confirming previous findings.

\subsubsection{\label{sec:stopping}Stopping}

Stopping in heavy-ion collisions has been of significant interest for a long time since it is of key importance for the initial energy transfer to the central rapidity region~\cite{Busza:1983rj,Bearden:2003hx,BRAHMS:2009fn}.
The study of stopping is here extended to the \pp system at 200 and 62.4\,GeV.   
 It has been known from  early low-energy \pp measurements \cite{PhysRevD.26.1497,AguilarBenitez:1991yy,PhysRevD.27.2580} that the fragmentation net-proton yields are nearly constant as a function of \xF. 
 In the case when the mean $m_T$ changes little vs.\ \xF this implies a \dndy distribution that  decreases exponentially with $y - y_{beam}$.

\begin{figure}[tb]
\centering
 {\includegraphics[width=3.2in]{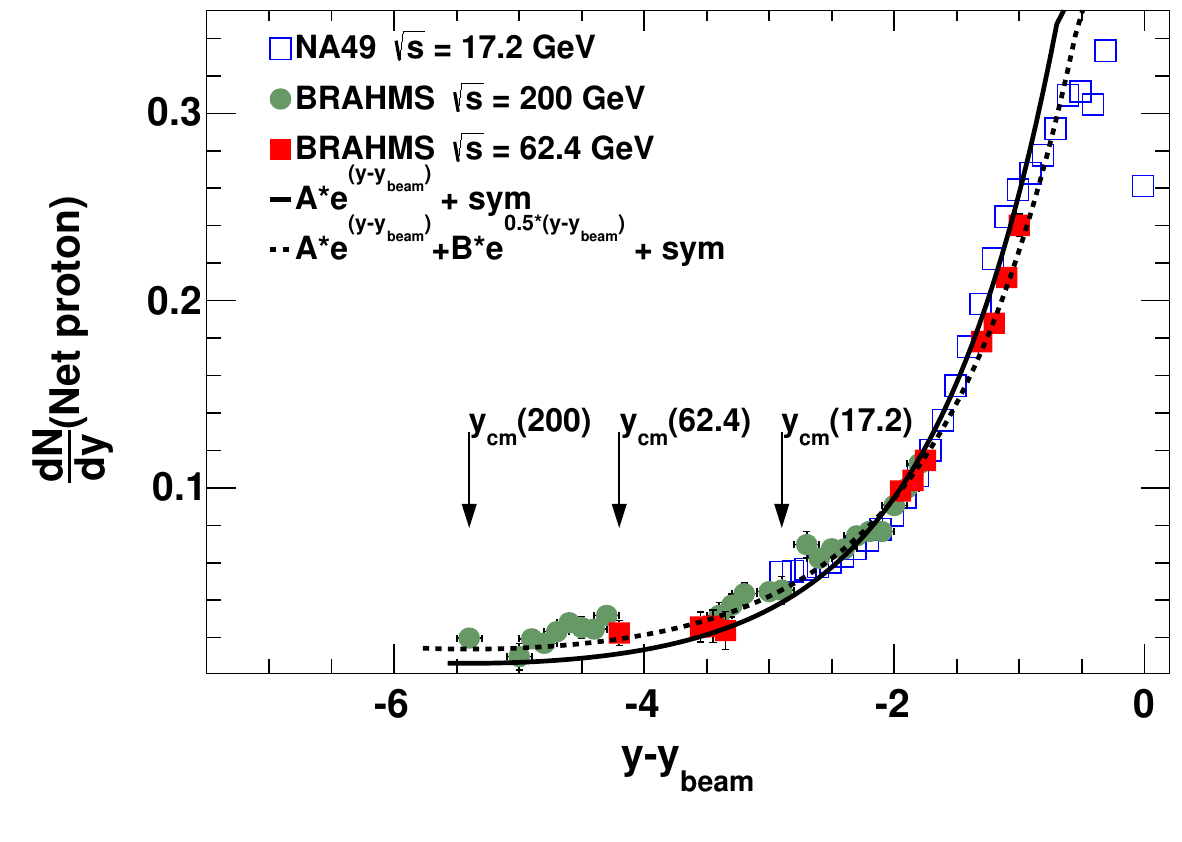}}
 \caption{\label{fig:netpscaling} 
 Net-protons from $\sqrt{s} = 200$, 62.4, and 17.2\,GeV (NA49). The data are 
 plotted to illustrate the longitudinal scaling. The three arrows
  indicate mid-rapidity, starting from the left, for 200, 62.4, and 17.2\,GeV \pp systems, respectively. 
  The solid  curve is an exponential with slope -1 ($e^{-y}$) symmetrized for the 200 GeV data to account for the symmetry around mid-rapidity, while  the dashed curve having two components with slopes $-1$ and $-0.5$ as discussed in the text.
  }
\end{figure}

In most experiments only charged hadrons are measured, so the net-proton distribution can be taken as a proxy for the net-baryon distribution.
 Figure~\ref{fig:netpscaling} shows the net-proton distribution plotted vs.\ $y - y_{beam}$. 
The net-proton values are derived from the difference of the proton and antiproton distributions in Fig.~\ref{fig:dndy_200_62}.
The data for net-proton \dndy\ from different beam energies have very similar values when compared at the same $y - y_{beam}$, thus no significant energy dependence of the stopping is observed from 17.2 to 200 GeV.
The mean rapidity loss, $\delta y$, is mainly determined by the first two units of rapidity loss.
The expected behavior from the lower energies has a dependence of $\exp{(-y)}$, which is shown as the solid curve in Fig.~\ref{fig:netpscaling}.
It describes the data quite well overall, but seems to underpredict the data at the highest values of $y - y_{beam} \sim -4$ to  -5.4.

As the rapidity range covered by combining all measurements is large, it is of interest to explore whether the present data may contain hints of effects from transport of protons via the baryon-junction mechanism as proposed by D.\ Kharzeev \cite{KHARZEEV1996238}. Recent heavy-ion data from STAR in Ref.~\cite{Lewis_2024} show that the net-proton measurements  at mid-rapidity vs. collision energy can be described with a rapidity shift of the form $\exp(-\alpha y)$, with the coefficient $\alpha = 0.61 \pm 0.03$. 
The slope $\alpha$ is approximately 1 at lower energies in \pp collisions  (up to rapidity losses of about 2). Since this behavior is well established, a single exponential dependence is unlikely to describe the full rapidity range.
In a recent paper \cite{Frenklakh:2023pwy}, a two-component picture consistent with a combined rapidity dependence of $\exp(-y)$ and $\exp(-0.5y)$ is proposed as a signature of baryon-junction transport (without arguing for a specific combination). 
In Fig.~\ref{fig:netpscaling}, we show with a dashed curve that an empirical sum of these two terms can give a fairly good description of the combined data from all three energies. 
The weights for the two exponential term were found by a fitting procedure involving a grid search.
The weight of the $\exp(-y)$ component is driven by the two lower energy data sets while the weight of the $\exp(-0.5y)$ is driven by the 200 GeV data.
The weights for the two components are 0.5 and 0.07.
Unfortunately, no measurements of the net-proton yield exist at even higher energies. 
The ALICE collaboration has measured the ratio of $\bar{p}/p$ at LHC energies at mid-rapidity in Refs.~\cite{PhysRevLett.105.072002,Alice2013} and those data are consistent with a long-range transport of protons to mid-rapidity  requiring a  slope $\alpha$ less than one.
As a concluding remark on this topic, we point out that an exponential decay with an $\alpha = 0.61$ exponent 
and anchored to mid-rapidity data, 
will be indistinguishable from the  sum of two exponentials (dashed curve), except for the forward region  $y - y_{beam} > -1.5$ where it underestimates data significantly.

Stopping is quantified by using the average rapidity loss,
$\mean{\delta y} = y_{beam} - \mean{y}$, where \mean{y} is calculated using \cite{PhysRevC.52.2684}
\begin{equation}
\mean{y} = \int_0^{y_{beam}}y\frac{dN_{B-\bar{B}}}{dy}dy.
\label{eq:averagey}
\end{equation}

In order to obtain the net-baryon \dndy from the measured net-proton 
\dndy, we make a correction according to Eq. \ref{stopping:dndy}
\begin{equation}
\frac{dN_{B-\bar{B}}}{dy} = \frac{dN_{p-\bar{p},meas}}{dy}\frac{n_p + n_n + n_\Lambda}{n_p + c_1n_\Lambda},
\label{stopping:dndy}
\end{equation}
where $\frac{dN_{p-\bar{p},meas}}{dy}$ is the number of measured net 
protons, $n_p$ is the number of true net-protons, $n_n$ is the number 
of net neutrons, and $n_\Lambda$ is the number of net $\Lambda$. The constant $c_1$ is the fraction of decay protons per $\Lambda$, registered  as primary protons. 
This procedure was used in Ref.~\cite{Bearden:2003hx}, where the constant was found to be $c_1 = 0.53\pm0.05$, with only a mild rapidity dependence, using Monte Carlo simulations~\cite{Bearden:2003hx}. 
This corresponds to approximately $83\%$ of the protons arising  from $\Lambda$ decays ($BR = 64\%$). This value of $c_1$ makes the conversion from net-protons to net-baryons surprisingly robust for the situation when the isospin is mainly carried by pions ($n_n \sim n_p$), where it is easy to see that Eq.~\ref{stopping:dndy} reduces to $dN_{B-\bar{B}}/dy \sim 2\,dN_{p-\bar{p},meas}/dy$.

The correction factor in Eq. \ref{stopping:dndy} can be rewritten as
\begin{equation}
\frac{1+\frac{n_n}{n_p}+\frac{n_\Lambda}{n_p}}{1+c_1\frac{n_\Lambda}{n_p}}
\label{stopping:dndycorrection}
\end{equation}
indicating that the important parameters are the net neutron to net 
proton ratio, $n_n/n_p$, and the net $\Lambda$ to net-proton ratio, $n_\Lambda/n_p$. 
\pythia is used to estimate $n_n/n_p$ and $n_\Lambda/n_p$ as a function of rapidity for \sqrts~ = 17.2, 62.4 and 200\,GeV. 
To constrain the ratios extracted from \pythia, the $\Lambda$ \dndy as measured by STAR~\cite{Abelev:2006cs} at \sqrts~ = 200 GeV  was combined with the BRAHMS proton measurement to obtain $n_\Lambda/n_p = 0.26\pm0.1$ at 
mid-rapidity. 
Fits to the \pythia predictions of $n_n/n_p$ and $n_\Lambda/n_p$ as a function of rapidity were used to obtain correction factors for each energy.
The corrections used in this study assume that \pythia predicts correctly 
the shape of the rapidity dependence of $n_n/n_p$ and $n_\Lambda/n_p$ at both energies
but ties the ratio to the experimental measurement of $n_\Lambda/n_p$ at mid-rapidity at \sqrts = 200\,GeV. 
The rapidity dependence for \pythia is to first order flat with a correction factor of 1.85 up to rapidity of 4.5 for 200 GeV and similar at 62.4 GeV up to 3.5. 
The EPOS event generator agrees with the experimental $\Lambda/p$ results at $y \sim 0$, but shows
a stronger  rapidity dependence for this ratio. Still, it results in a similar mean correction factor.
\begin{figure}[tb]
\centering
 {\includegraphics[width=3.2in,angle=0]{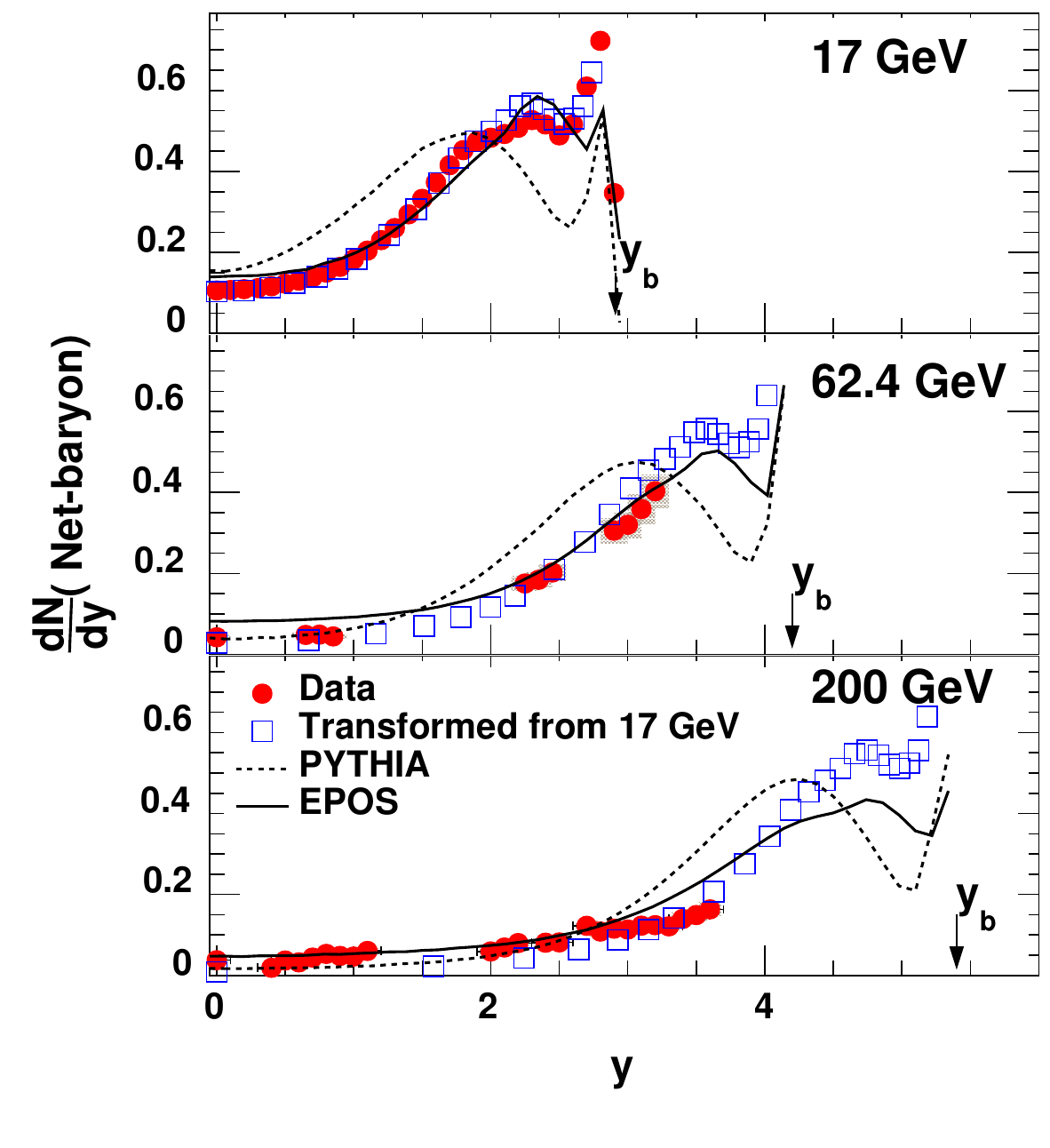}}
 \caption{\label{fig:stopping} 
Net-baryon \dndy\ distribution for 62.4 GeV (center panel) and 
200 GeV (bottom panel). The data are indicated by the solid points 
and the predictions of the \pythia and \epos calculations are shown by the solid and dashed curves, respectively. 
The open squares in the center and bottom panel show the net-baryon $dN/dx_{\rm F}$ at 17.2\,GeV transformed, assuming $x_{\rm F}$ scaling, to \dndy\ at the respective energies. The top panel shows the same net-baryon $dN/dx_{\rm F}$ at 17.2\,GeV transformed to \dndy\ as red data points and the \dndy\ distribution derived from the 17.2\,GeV net-proton distribution~\cite{Anticic:2009wd} with blue open squares. 
 }
\end{figure}

Using these factors in Eq.~\ref{stopping:dndycorrection}, one obtains the net-baryon 
rapidity distribution for the data at both RHIC energies, which are shown as solid circles in Fig.~\ref{fig:stopping}. 
The top panel shows the net-baryon rapidity distribution derived from the NA49 data~\cite{Anticic:2009wd}. 
The solid  (\pythia) and dashed (\epos) curves will be discussed in Section~\ref{sec:theory}.

The 200 and 62.4\,GeV data in Fig.~\ref{fig:stopping} show a plateau of very low $\frac{dN_{B-\bar{B}}}{dy}$ near mid-rapidity followed by an increase as the beam rapidity is approached, reaching higher values within the BRAHMS acceptance for the 62.4\,GeV data than for the 200\,GeV data. 
The 17.2 GeV data~\cite{Anticic:2009wd} have full acceptance from mid-rapidity to the beam rapidity and the good agreement observed at similar values of $y - y_{beam}$ for different beam energies suggests  that these  data can be used to constrain the measurements presented here in the regions where there is no experimental acceptance.
The $dN/dx_{\rm F}$ data from 
Ref.~\cite{Anticic:2009wd} are used to transform into \dndy at higher energies. 
Figure \ref{fig:stopping} shows the results of this transformation, indicated with open blue squares, compared with the BRAHMS measurements. 
The agreement between the net-baryon rapidity distributions measured by BRAHMS at 62.4 and 200 GeV and the transformed SPS measurement is remarkably good once the correction factors defined with Eq.~\ref{stopping:dndycorrection} are applied. 
These agreements again indicate that \mean{\delta y} does not change substantially over the energy range from 
17 GeV to 200 GeV.

\subsection{Average multiplicities versus energy}
\label{sec:mult}
 Reference \cite{Antinucci:1972ib} contains a summary of average identified particle multiplicities, integrated over all phase space, up to and including ISR energy data. 
The present BRAHMS data from \pp at $\sqrt{s}=200$ and 62.4 GeV provide additional entries to the summary by supplying measurements of \dndy up to about rapidity 3.5-4 for charged hadrons. 
The $4\pi$ yields are estimated by two means: 
first, by fitting the rapidity distributions to a Gaussian distribution and extracting the yields from the functional integrals, and,  second by integrating over the measured range and extrapolating linearly to $y_{beam}$ or slightly above assuming \dndy\ is zero at the limit.
The measurements typically cover more than $85\%$ of the $4\pi$ yield assuming symmetry around $y=0$. At 200 GeV the extrapolation accounts for 10-15\%  for pions and 5-7\% for \pbar. At 62.4 GeV the extrapolations account for less than 5\% of the integrated yields.
The systematic uncertainties are estimated taking into account the uncertainties from the normalization of mid-rapidity and forward rapidity data, respectively.
The data for 200 GeV are presented in Table~\ref{tab:datamodel200}.
 It tabulates for each hadron species the integrated yield, the deduced RMS as well as results from \pythia and \epos calculations to be discussed later. 
The largest contribution to the systematic uncertainty comes from the absolute normalization which is correlated among the different particle types and therefore not included for the individual $4\pi$ yields.
 The corresponding data for 62.4 GeV are given in Table~\ref{tab:datamodel62}.

\begin{table*}[htb]
\begin{tabular}{|c|c|c|c|c|c|c|}
\hline
200 GeV&Data&&\epos&&\pythia&\\
\hline
Particle&\mean{N}&RMS&\mean{N}&RMS&\mean{N}&RMS\\
\hline
$\pi^{+}$  &$8.92\pm0.28\pm1.18$ & $2.43\pm0.15$&8.31&2.48&8.50&2.47  \\
\hline
$\pi^{-}$ &$8.48\pm0.28\pm1.13$& $2.36\pm0.15$&7.65&2.32&7.77&2.31\\
\hline
\kplus&
$0.84\pm0.04\pm0.11$&
$2.30\pm0.15$&
0.88&2.26&0.80&2.19\\
\hline
\kminus&
$0.69\pm0.06\pm0.09$&
$2.14\pm0.15$&
0.75&
2.06&
0.70&2.00\\
\hline
\pbar&
$0.41\pm0.02\pm0.05$&
$1.86\pm0.15$&
0.40&1.76&
0.42&1.86\\

\hline

\end{tabular}

\caption{\label{tab:datamodel200}
Integrated $4\pi$ yields and the RMS for \dndy distributions at 200 GeV with the corresponding model results for \epos and \pythia. The quoted uncertainties for the yields are statistical and systematic respectively. Additional, there is a normalization uncertainty of 12.3\%, which is not included. The uncertainty on the RMS includes both statistical and   systematic uncertainties. }
\end{table*}


\begin{table*}[htb]
\begin{tabular}{|c|c|c|c|c|c|c|}
\hline
62.4 GeV&Data&&\epos&&\pythia&\\
\hline
Particle&\mean{N}&RMS&\mean{N}&RMS&\mean{N}&RMS\\
\hline
$\piplus$  &$4.79\pm0.20\pm0.77$&
$2.04\pm0.2$&
5.28&2.02&5.70&1.98  \\
\hline
$\pi^{-}$ &$3.94\pm0.16\pm0.63$& $1.89\pm0.2$&
4.61&1.85&4.96&1.84\\
\hline
\kplus&
$0.37\pm0.02\pm0.06$&
$1.84\pm0.15$&
0.50&1.78&0.51&1.70\\
\hline
\kminus&
$0.27\pm0.02\pm0.04$&
$1.58\pm0.15$&
0.38&
1.59&
0.42&1.51\\
\hline
\pbar&
$0.15\pm0.02\pm0.02$&
$1.30\pm0.15$&
0.16&1.33&
0.27&1.35\\

\hline

\end{tabular}

\caption{\label{tab:datamodel62}
Integrated $4\pi$ yields and the RMS for \dndy distributions at 62.4 GeV with the corresponding model results for \epos and \pythia. The quoted uncertainties for the yields are statistical and systematic respectively. Additional, there is a normalization uncertainty of 11.7\%, which is not included. The uncertainty on the RMS includes both statistical and   systematic uncertainties.}
\end{table*}

From the rapidity distributions we also extract the RMS value assuming that they are symmetric around $y=0$ and by using the same extrapolation to $y_{beam}$ as for the $4\pi$ integrated yields. 
For the 62.4 GeV data the RMS is essentially determined from the data in the measured range.
At 200 GeV the extrapolation uncertainty contributes about 0.1 unit. 
The overall features are that the RMS for \piplus\ and \piminus\ are similar, with the \kplus\ and \kminus\ being slightly narrower, and with quite narrow  \pbar\ distributions.

The charged-particle $4\pi$~multiplicity is obtained by summing up the pion, kaon and antiproton multiplicities and adding the beam fragmentation protons, where 
the latter amounts to ${\approx}1.2$ protons per collision, independent of beam energy \cite{PhysRevC.52.2684}. 
As the normalization systematic is common for all the measured particles it is excluded from the quoted systematic uncertainty and given separately. 
The total charged-particle multiplicity extracted from the BRAHMS 200 GeV data, i.e., the sum of charged pions, kaons, and protons is $21.0\pm 0.41 (\text{stat}) \pm 2.78 (\text{syst}) \pm 2.68 (\text {norm})$. This is in good agreement with the PHOBOS charged-hadron $dN/d\eta$ measurement of $20.2\pm 1.8$ \cite{Alver:2010ck}.
The present 62.4 GeV $4\pi$-integrals  of identified particles seem slightly   lower, in particular for kaons, than the overall systematics as extracted from the ISR measurements. 
The total 4$\pi$ charge-particle multiplicity is $10.9\pm 0.26(\text{stat}) \pm 0.78                                                                  (\text{syst}) \pm 1.27 (\text {norm})$ in good agreement with the value of $12.26\pm0.21$ for the inelastic cross section of Ref.~\cite{Breakstone:1983ns}.
Total $4\pi$ charged-particle multiplicity results are not available from 62.4 GeV ISR data, but the 53 GeV data from Ref.~\cite{Rossi:1974if} quotes $4.68\pm 0.47, 4.29\pm 0.43, 0.43\pm 0.06, 0.31\pm0.05$, and $0.14\pm 0.02$ for \piplus, \piminus, \kplus, \kminus, and \pbar, respectively, per inelastic collision.
These are slightly higher than the BRAHMS results, considering the expected additional yield going from 53 to 62.4 GeV.
It should be emphasized that the integral $4\pi$ values from ISR for identified particles have been derived from measurements at
 \pt= 0.4 \gevc\ and assume a rapidity-independent \pt spectrum shape ($e^{-B \pt}$),
 which based on the BRAHMS measurements, is not warranted up to the beam rapidity $y\sim 4.2$.
 If we use such parametrization in extracting the 
 \dndy yields from the present data, the forward rapidity cross sections for kaons increase by 20-40\%, but with a considerably worse $\chi^{2}$  of the spectral fits. 
 It should be reiterated that the \dndy at $y=0$ is in good agreement with both ISR and PHENIX measurements.
The only previous measurement of $4\pi$ yields at 200 GeV is for $K^{0}_s$, with \mean{N } = $0.72\pm 0.12$\cite{Ansorge:1987cj}, which is in good agreement with the present $K^{+}$ and $K^{-}$ multiplicities.

\section{\label{sec:theory}Comparison with \pythia and \epos}

This section compares selected minimum bias (inelastic) results with calculations using \pythia and \epos.
The focus of the comparison with data is on soft QCD observables, such as  \pt-spectra vs.\ rapidity and \dndy for identified charged hadrons. 
The comparison with the net-proton distributions is of particular interest.

In \pythia~\cite{PYTHIA,Bierlich:2022pfr}, all inelastic non-diffractive interactions are modeled as a sum of perturbative QCD parton-parton interactions, with soft processes dominated by low-momentum transfers. The Lund string model~\cite{Andersson:1983ia} is used to model how partons hadronize. The \pythia calculations are done using version 8.226 and the default tune~\cite{Corke:2010yf,Corke:2010zj}. This tune describes Tevatron and mid-rapidity RHIC high-\pt data quite well, as well as charged particle data from LHC \cite{Corke:2010yf}. 

The \epos model, described in detail in Ref.~\cite{Werner:2005jf}, uses a multiple scattering approach based on parton ladders to describe the initial interactions. It has two widely different ways to handle the final-state partons: dilute ``corona'' regions handled with strings while dense ``core'' regions are described as thermal quark-gluon plasmas.
\epos has had considerable success in describing both hadronic and heavy-ion processes, e.g., early RHIC \pp and d+Au data, and has also been applied extensively to explain very forward Cosmic Ray observations.
 In this paper we use the \epos 1.99 version as described in Ref.~\cite{Pierog:2009zt} with its default settings. 
 
The BRAHMS transverse momentum spectra and rapidity density 
distributions are normalized to the total inelastic cross section, which means that NSD, Single and Double Diffractive (SD/DD) processes are included in the calculations (but not elastic processes). 

\begin{figure*}[tbp]
 \centering
 {\includegraphics[width=\linewidth]{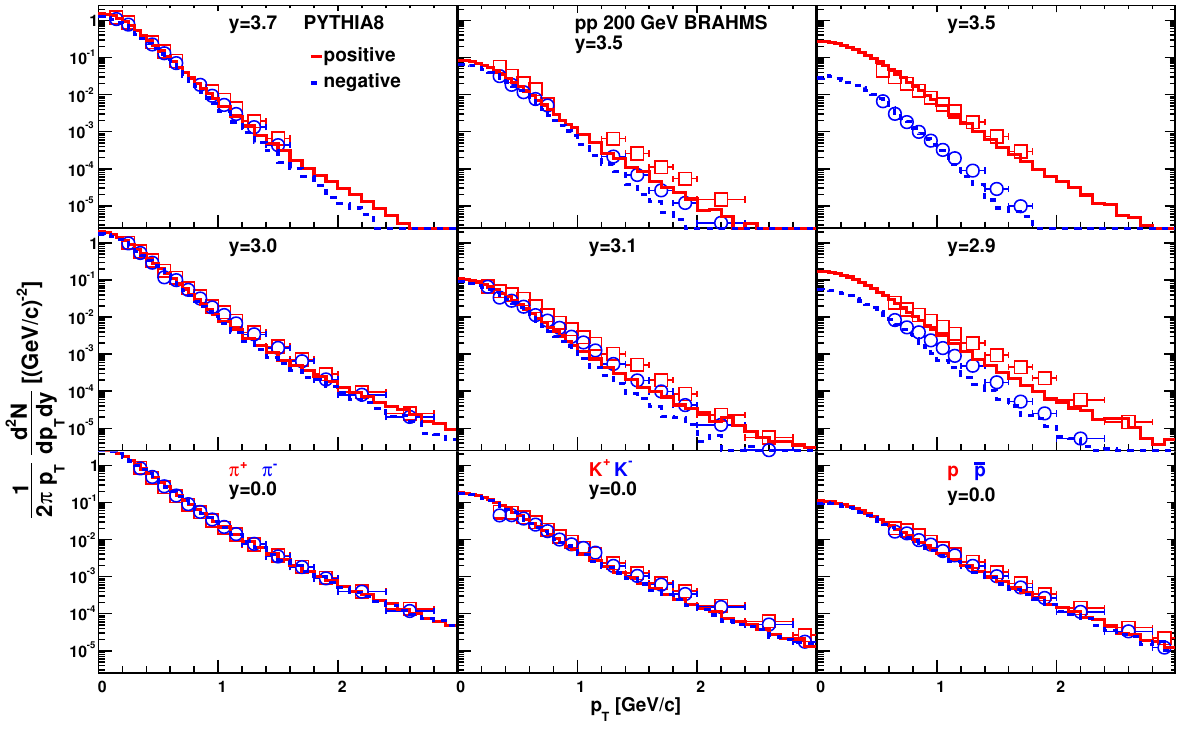}}
 \caption{\label{fig:pythia200all} 
  Invariant transverse spectra for \piplus, \kplus, \kminus, p and 
\pbar at $y=0$
 and at two forward rapidities, as indicated in the figure, compared with 
\pythia calculations.
For kaons and protons, the positive particles are indicated by 
(red) squares and for negative particles by (blue) circles. The data are from 200 GeV \pp collisions.
 }
\end{figure*}

\begin{figure*}[tbp]
 \centering
 {\includegraphics[width=\linewidth]{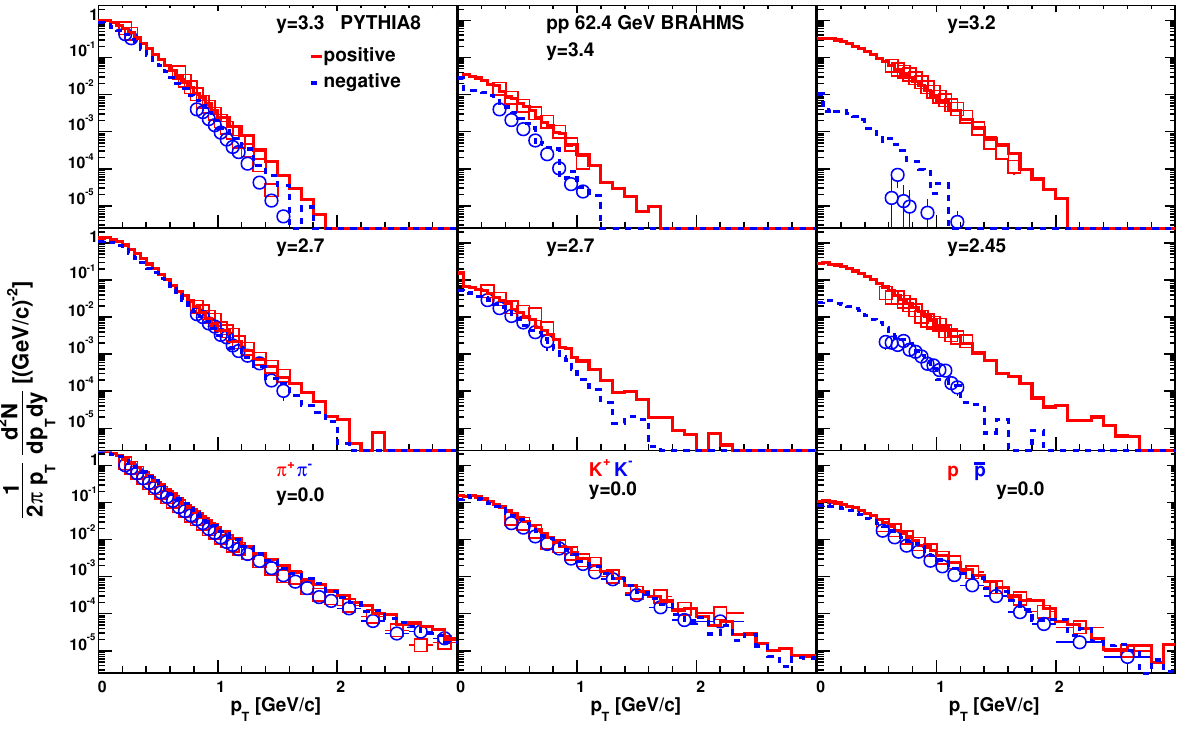}}
 \caption{\label{fig:pythia62all} 
  Invariant transverse spectra for \piplus, \piminus, \kplus, \kminus, p and \pbar at y=0
 and at two forward rapidities, as indicated in the figure, compared with \pythia calculations.
 Positive particles are indicated by (red) squares and negative particles by
 (blue) circles. The data are from 62.4 GeV \pp collisions.
 }
\end{figure*}

Figures \ref{fig:pythia200all} and \ref{fig:pythia62all} show  transverse momenta spectra at mid-rapidity and at two forward rapidities for $\sqrt{s} = 200$\,GeV and $\sqrt{s} = 62.4$\,GeV, respectively. The solid (red) and dashed (blue) curves show the \pythia calculations for positive and negative hadrons, respectively. The qualitative agreement is good for all particles at both energies and the model describes the softening of the \pt-spectra for all species as rapidity increases.
The splitting between the favored (positive) and unfavored (negative) particles with increasing rapidity is also reproduced.
At the selected rapidities the shape of the proton and antiproton spectra are in general qualitatively well described, but there is significant tension for protons at forward rapidity and, as is shown later, the rapidity distributions are therefore not well captured by \pythia. 

\begin{figure*}[tbh]
 \centering
 {\includegraphics[width=\linewidth]{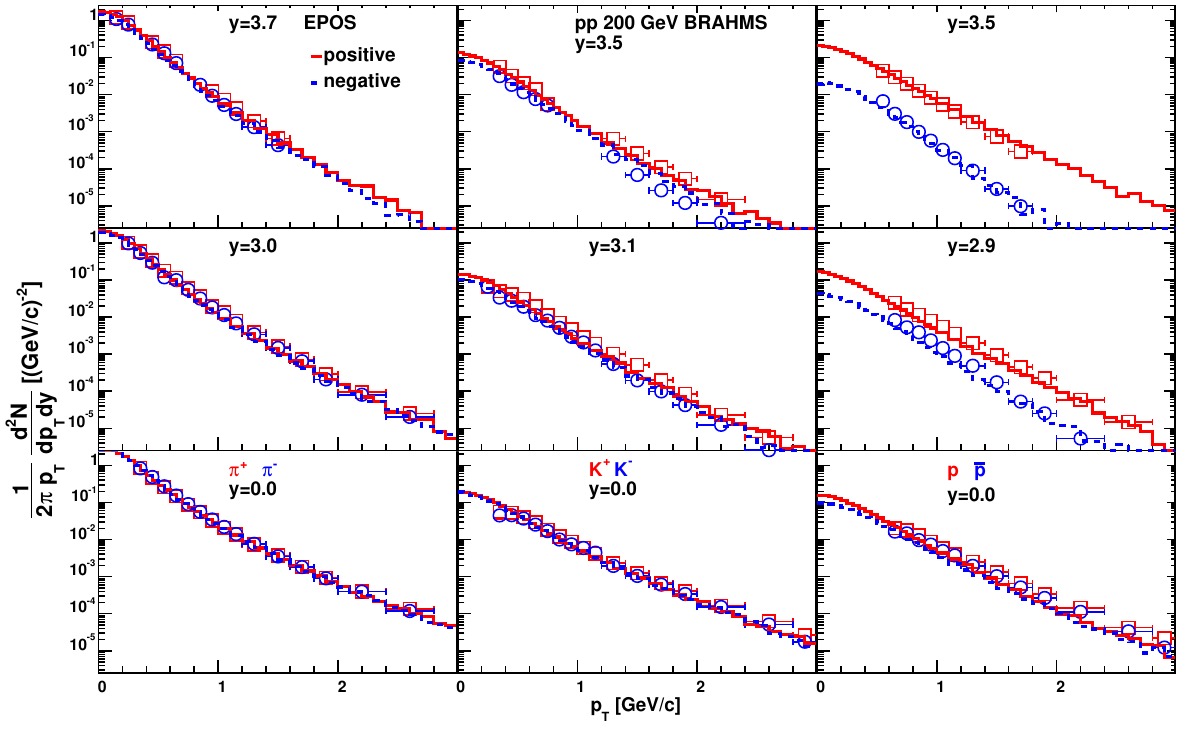}}
 \caption{\label{fig:epos200all}  
  Invariant transverse spectra for \piplus, \piminus,  \kplus, \kminus, p and 
\pbar at $y=0$
 and at two forward rapidities, as indicated in the figure, compared with \epos calculations.
For kaons and protons the positive particles are indicated by the 
squares (red) and by the open circles (blue) for the negatives. The data are from  200 GeV \pp collisions.}
\end{figure*}

\begin{figure*}[tbh]
 \centering
 {\includegraphics[width=\linewidth]{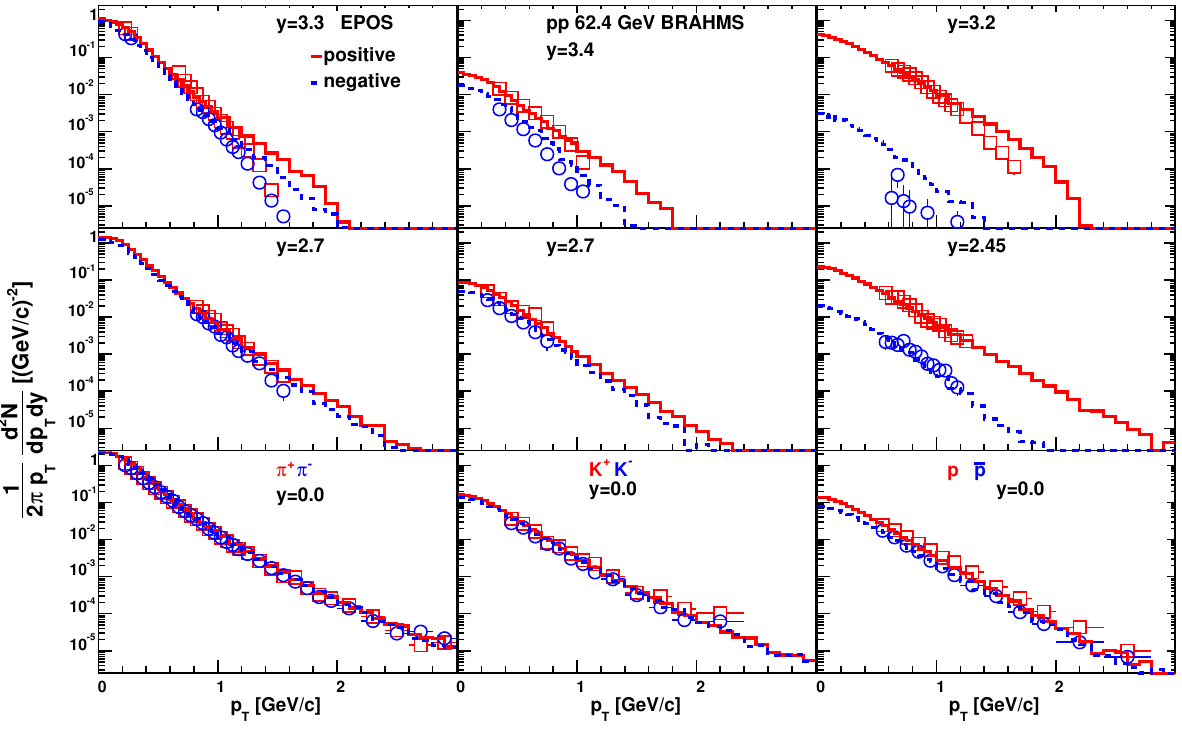}}
 \caption{\label{fig:epos62all} 
  Invariant transverse spectra for \piplus, \piminus,  \kplus, \kminus, p and 
\pbar at $y=0$
 and at two forward rapidities, as indicated in the figure, compared with \epos calculations.
For kaons and protons the positive particles are indicated by the 
squares (red) and by the open circles (blue) for the negatives. The data are from  62.4  GeV \pp collisions.}
\end{figure*}
In Fig.~\ref{fig:epos200all} and \ref{fig:epos62all} the same \pt spectra are compared with \epos calculations. The agreement is good for pions at both energies, but only qualitative agreement is found for kaons and protons. Looking back at Fig.~\ref{fig:meanptall}, one can see that the \meanpt\ for the kaons and protons is underpredicted, particularly at mid-rapidity and at 200 GeV. 
The same trend has been noted for \pythia but not shown in Fig.~\ref{fig:meanptall}.

\begin{figure}[tb]
 \centering
 {\includegraphics[width=3.2in]{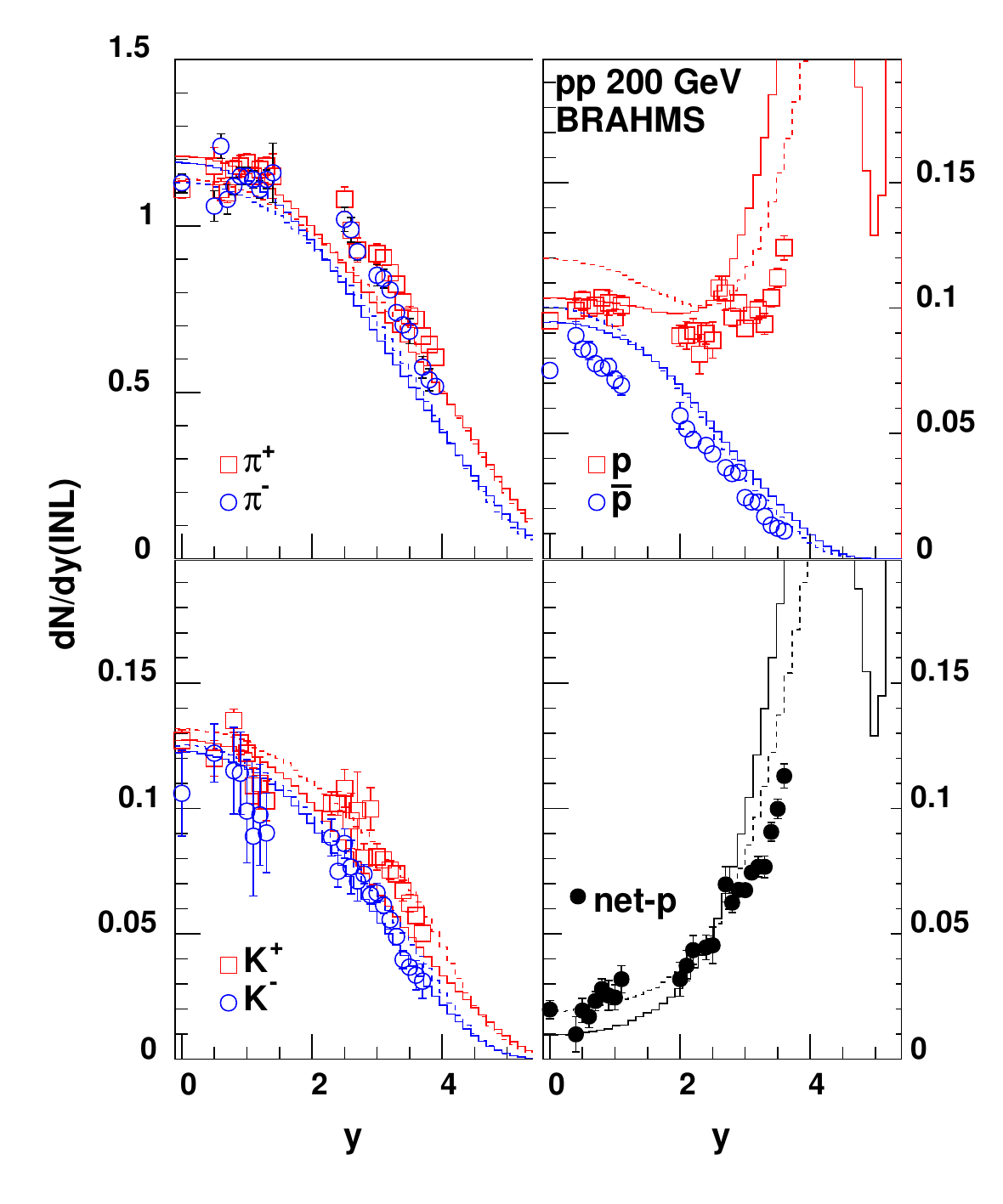}}
 \caption{\label{fig:dndy_200} 
  Rapidity distribution for \piplus, \kplus, \kminus, p and \pbar and net-proton at 200 GeV compared with {\pythia} (solid lines) and \epos (dashed) calculations.
  The positive and negative particles are indicated by the red squares 
and the blue circles, respectively.
}
\end{figure}

\begin{figure}[tb]
 \centering
 {\includegraphics[width=3.2in]{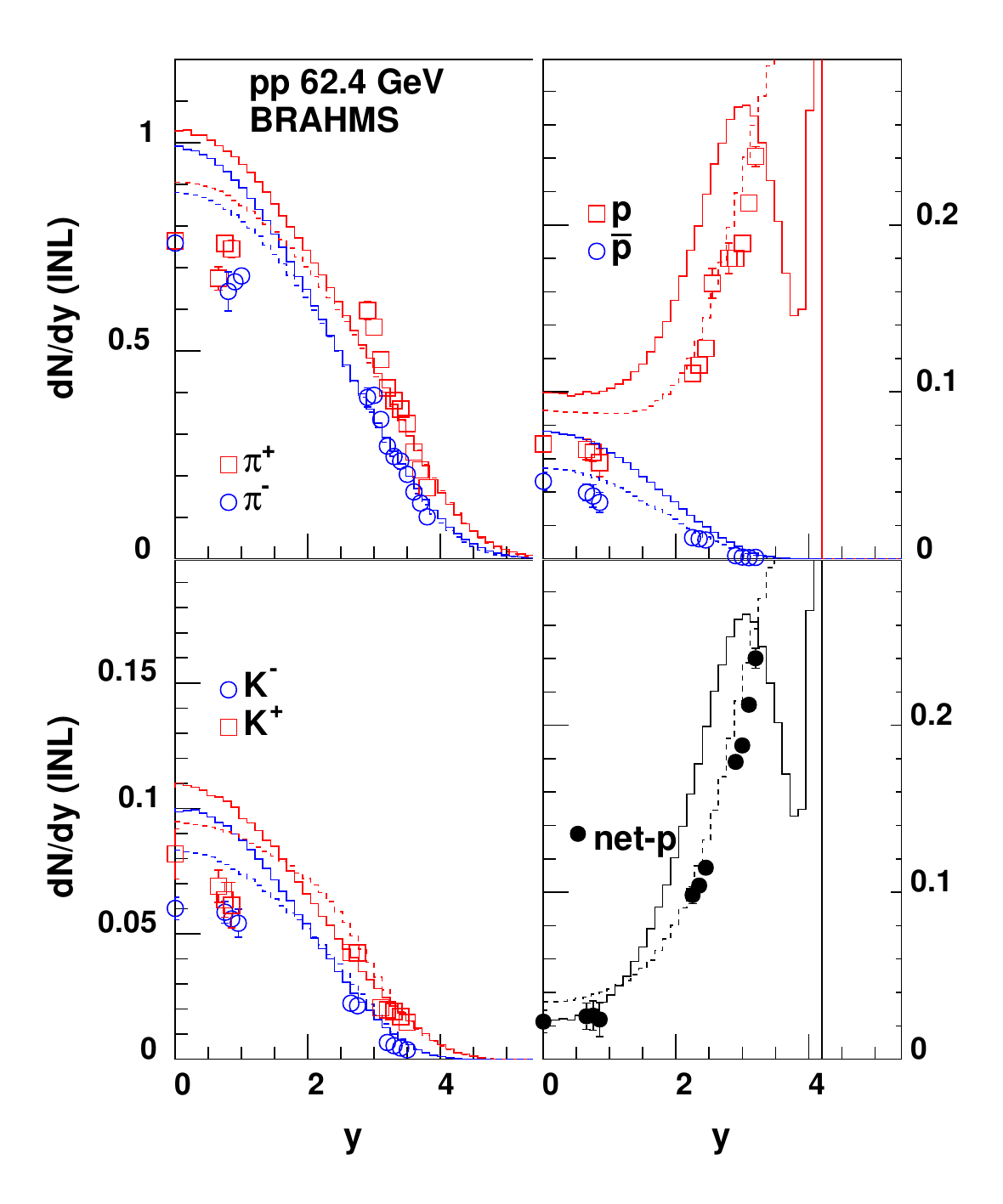}}
 \caption{\label{fig:dndy_62} 
  Rapidity distribution for \piplus, \kplus, \kminus, p and \pbar and net-proton at 62.4 GeV compared with {\pythia} (solid lines) and \epos (dashed lines) calculations.
 The positive and negative particles are indicated by the red squares 
and the blue circles, respectively.
 }
\end{figure}

The model comparisons with \dndy\ as a function of rapidity are shown in Fig.~\ref{fig:dndy_200} for 200 GeV and in Fig.~\ref{fig:dndy_62} for the 62.4 GeV data. For {\pythia} there is reasonable agreement for pions, kaons and antiprotons at 200 GeV, whereas at  62.4\,GeV 
all species are somewhat overpredicted by the model calculations.
For \epos, the agreement for pions and kaons is in general better than for \pythia, in particular for the 62.4\,GeV mid-rapidity results, where \pythia significantly overpredicts the meson yields. 
Where \epos stands out is that the proton and net-proton rapidity distributions are much better described than by \pythia,  most strikingly at forward rapidity.
A similar conclusion was reached by NA61~\cite{Aduszkiewicz:2017sei}, which measured $\pi$, K,  and p in \pp collisions at SPS energies where the EPOS model describes all rapidity density distributions very well.
The baryon stopping is well described by EPOS over a wide energy range, in contrast to many other models.
The $4\pi$ integrated yields and the RMS of the rapidity distributions have also been obtained for \pythia and \epos and are included in Tabs.\ \ref{tab:datamodel200} and \ref{tab:datamodel62}.
For the model comparison regarding stopping we return to  Fig.~\ref{fig:stopping} where the $B-\bar{B}$ 
rapidity distribution obtained for \pythia and \epos are compared with the new BRAHMS results as well as the NA49 measurement. 
The most glaring discrepancy between \pythia and the data appears to be the failure in producing the proper rapidity distributions of net-baryons, as derived from the net-protons data.
This points to issues in either the modeling of processes involving the valence quarks in the projectile and
target and/or the hadronization of the beam remnants.
In contrast, it appears that \epos provides a better description of the fragmentation region, in particular for the baryons.  

\section{Summary}

The BRAHMS experiment at RHIC has performed measurements of charged pions, kaons, and (anti)protons in \pp collisions at \sqrts~= 62.4 and 200\,GeV over the most extended rapidity interval  to date. 
The BRAHMS \pt spectra at mid-rapidity agree well with data from ISR and PHENIX at 62.4 GeV, and with STAR data at 200 GeV, but are higher than the PHENIX data at 200 GeV.
From the extrapolated rapidity densities, the RMS and integrated multiplicities over the full $4\pi$ solid angle were obtained for the first time at these beam energies.

Model comparisons show that {\pythia} describes the pion and kaon spectra reasonably well at both energies and slightly overpredicts the yields at 62.4 GeV. \epos provides an overall better description of the data, particularly for protons and net-protons, where PYTHIA shows significant deviations. Both models underpredict the \meanpt for kaons and protons.

Longitudinal scaling is observed for net-proton \dndy\ distributions from $\sqrts = 17\,\text{GeV}$ to 200\,GeV. 
The observed scaling qualitatively agrees with a simple exponential decrease from beam rapidity and is approximately consistent with a flat distribution in $dN/dx_F$, which was observed at lower beam energies.
However, we find that the net-proton rapidity yields are better described when an additional long-range exponential component is included, 
which may be an indication of a contribution  of baryon junction transport from beam- to mid-rapidity at high beam energies.

\acknowledgments 

We thank the staff of the Collider-Accelerator, the Physics Department and SDCC at BNL for their vital contributions. 
This work was supported by the office of Nuclear Physics of the
within Office of Science of the U.S. Department of Energy, the Danish Natural Science Research Council,
the Research Council of Norway, 
the Polish Ministry of Science and Higher Education (Contract no 1248/B/H03/2009/36), 
and the Romanian Ministry of Education and Research for the
grant no.81-049/2007 (REEHUC)
and a sponsored research grant from Renaissance Technologies Corp.

\clearpage

\bibliographystyle{JHEP}
\bibliography{pp_paper}
\end{document}